\documentclass[a4paper,11pt]{article}
\pdfoutput=1
\usepackage{jheppub}
\usepackage[utf8]{inputenc} 
\usepackage[english]{babel}
\usepackage{amsthm,amsfonts}
\usepackage{accents}
\usepackage{dutchcal}
\usepackage{upgreek}
\allowdisplaybreaks
\usepackage{lieart}
\usepackage{enumitem}
\usepackage{tikz}
\usetikzlibrary{arrows}
\usetikzlibrary{calc}
\tikzset{>=latex}
\usepackage{dsfont}
\usepackage{cite}
\usepackage{mcite}

\DeclareMathOperator{\diag}{diag}

\newcommand{\dd}{\mathrm{d}}

\newcommand{\PS}{Pol\'a\v{c}ek-Siegel}

\newcommand{\hrangle}{\raisebox{-3px}{\resizebox{6px}{12px}{$\succ$}}}
\newcommand{\hlangle}{\raisebox{-3px}{\resizebox{6px}{12px}{$\prec$}}}
\newcommand{\heta}{\upeta}
\newcommand{\htau}{\uptau}
\newcommand{\hatt}{\widehat{t}}
\newcommand{\Khet}{\mathcal{K}}
\newcommand{\Rhet}{\mathcal{R}}
\newcommand{\Rthet}{\widetilde{\mathcal{R}}}

\newcommand{\hRt}{\widetilde{\mathcal{R}}}
\newcommand{\hK}{\mathcal{K}}

\newcommand{\Pb}{\overline{P}}

\newcommand{\Odd}[1][d]{\mathrm{O}(#1,#1)}
\newcommand{\GL}[1][n]{\mathrm{GL}(#1)}
\newcommand{\GH}{G_{\mathrm{H}}}
\newcommand{\GS}{G_{\mathrm{S}}}

\newcommand{\GD}{G_{\mathrm{D}}}
\newcommand{\GM}{G_{\mathrm{M}}}
\newcommand{\GPS}{G_{\mathrm{PS}}}
\newcommand{\gPS}{\mathfrak{g}_{\mathrm{PS}}}

\newcommand{\gM}{\mathfrak{g}_{\mathrm{M}}}
\newcommand{\fp}{\mathfrak{p}}

\newcommand{\Rt}{\widetilde{R}}

\newcommand{\Ah}{\hat{A}}
\newcommand{\Bh}{\hat{B}}
\newcommand{\Ch}{\hat{C}}
\newcommand{\Dh}{\hat{D}}
\newcommand{\Eh}{\hat{E}}
\newcommand{\Fh}{\hat{F}}
\newcommand{\Phih}{\hat{\Phi}}
\newcommand{\Dt}{\widetilde{D}}
\newcommand{\Ih}{\hat{I}}
\newcommand{\Jh}{\hat{J}}
\newcommand{\Kh}{\hat{K}}
\newcommand{\Lh}{\hat{L}}

\newcommand{\Ac}{\mathcal{A}}
\newcommand{\Bc}{\mathcal{B}}
\newcommand{\Cc}{\mathcal{C}}
\newcommand{\Dc}{\mathcal{D}}
\newcommand{\Ec}{\cE}
\newcommand{\Fc}{\cF}
\newcommand{\Ic}{\mathcal{I}}
\newcommand{\Kc}{\mathcal{K}}

\newcommand{\Ad}{\mathds{A}}
\newcommand{\Bd}{\mathds{B}}
\newcommand{\Cd}{\mathds{C}}

\newcommand{\genLieM}{\widehat{\mathcal{L}}}
\newcommand{\genLieHet}{\mathcal{L}}
\newcommand{\genLie}{\mathbb{L}}
\newcommand{\Zop}{\mathbf{Z}}
\newcommand{\ZHet}{\mathcal{Z}}
\newcommand{\Cop}{\mathbf{C}}
\newcommand{\hE}{\widehat{E}}
\newcommand{\Mt}{\widetilde{M}}
\newcommand{\mt}{\widetilde{m}}
\newcommand{\cE}{\mathcal{E}}
\newcommand{\cA}{\mathcal{A}}

\newcommand{\cF}{\mathcal{F}}

\newcommand{\cH}{\mathcal{H}}
\newcommand{\bcF}{\boldsymbol{\cF}}

\newcommand{\Vt}{\widetilde{V}}
\newcommand{\Bt}{\widetilde{B}}
\newcommand{\Ht}{\widetilde{H}}

\newcommand{\Th}{\widehat{T}}
\newcommand{\Ti}{\mathring{T}}
\newcommand{\Te}{\check{T}}
\newcommand{\Thetai}{\mathring{\Theta}}
\newcommand{\THet}{\mathcal{T}}
\newcommand{\THeti}{\mathring{\mathcal{T}}}
\newcommand{\Thetab}{\overline{\Theta}}

\newcommand{\alphai}{\tilde{\alpha}}
\newcommand{\alphan}{\underaccent{\tilde}{\alpha}}
\newcommand{\betai}{\tilde{\beta}}
\newcommand{\betan}{\underaccent{\tilde}{\beta}}
\newcommand{\gammai}{\tilde{\gamma}}
\newcommand{\gamman}{\underaccent{\tilde}{\gamma}}
\newcommand{\deltai}{\tilde{\delta}}
\newcommand{\deltan}{\underaccent{\tilde}{\delta}}

\newcommand{\epsilonn}{\underaccent{\tilde}{\epsilon}}

\newcommand{\mui}{\tilde{\mu}}
\newcommand{\mun}{\underaccent{\tilde}{\mu}}
\newcommand{\onei}{\tilde{1}}

\newcommand{\aL}{\underline{a}}
\newcommand{\aR}{\overline{a}}
\newcommand{\bL}{\underline{b}}
\newcommand{\bR}{\overline{b}}
\newcommand{\cL}{\underline{c}}
\newcommand{\cR}{\overline{c}}

\newcommand{\oneL}{\underline{1}}
\newcommand{\oneR}{\overline{1}}

\newcommand{\xih}{\widehat{\xi}}

\newcommand{\llangle}{\langle\!\langle}
\newcommand{\rrangle}{\rangle\!\rangle}

\newsavebox\MBox

  \title{\boldmath Duality covariant curvatures for the heterotic string}%\preprint{}

\author[a]{Falk Hassler,}
\author[a]{David Osten,}
\author[b]{and Yuho Sakatani}

\emailAdd{falk.hassler@uwr.edu.pl}
\emailAdd{david.osten@uwr.edu.pl}
\emailAdd{yuho@koto.kpu-m.ac.jp}

\affiliation[a]{University of Wrocław, Faculty of Physics and Astronomy, Maksa Borna 9, 50-204 Wrocław, Poland}
\affiliation[b]{Department of Physics, Kyoto Prefectural University of Medicine,
1-5 Shimogamohangi-cho, Sakyo-ku, Kyoto 606-0823, Japan}

\abstract{Duality covariant curvature and torsion tensors in double field theory/generalized geometry are central in analyzing consistent truncations, generalized dualities, and related integrable $\sigma$-models. They are constructed systematically with the help of a larger, auxiliary space in a procedure inspired by Cartan geometry originally proposed by Pol\'a\v{c}ek and Siegel for bosonic strings. It pivots around a maximally isotropic group that captures the generalized structure group of the physical space. We show how dropping the isotropy condition on this group allows us to describe heterotic/type I strings. As an immediate application, we construct a new family of heterotic backgrounds that interpolates between the two-dimensional cigar and trumpet backgrounds.}

\begin{document}

\maketitle

\section{Introduction}
Generalized geometry and the closely related double/exceptional field theories provide a unified framework to analyze the dynamics of strings, membranes and their low-energy effective supergravity limit. They extend fundamental objects from differential geometry and seek a physical interpretation for them. Important examples are the generalized metric and the related generalized frame. Like the metric in general relativity, they aim to unify all physical fields of more complicated theory with gravity. Depending on the implemented duality group, these fields include besides the metric various form fields and even fermions for supergroups. In contrast to general relativity, it has proven very difficult to define a curvature tensor in this setup. While a generalized version of the torsion tensor can be constructed from the generalized Lie derivative, the generalized Riemann tensor is more elusive. A first naive guess -- the Riemann tensor of standard differential geometry -- does not transform covariantly. It has to be further modified and projected. While these steps eventually provides a result, its geometric interpretation is not clear and a more natural way to define curvature is desirable. One might argue that this is a purely mathematical problem because to describe a physical system an action which is invariant under all symmetries of the theory is sufficient. However, recently it became clear that a more refined understanding of curvatures in generalized geometry is central in the exploration of dualities and consistent truncations \cite{Butter:2022iza,Hassler:2023axp}.

In this regard a construction initially proposed by Pol\'a\v{c}ek and Siegel in \cite{Polacek:2013nla} has proven to be very helpful. Instead of working directly with the generalized metric, it rather employs a frame formalism to permit a straightforward supersymmetric extension. One encounters two relevant symmetries in this approach: generalized diffeomorphisms that unify diffeomorphisms with gauge transformations (and even supersymmetry transformations for super-duality groups), and generalized Lorentz transformations -- encoding the transformation of fermions. A curvature tensor should transform covariantly under both of them. What we now call the \PS{} construction achieves this objective for O($d$,$d$) generalized geometry by starting from an extended space. This in particular requires to introduce additional connections besides generalized geometry's analog of the spin connection. Recently, this idea has been refined in various ways to
\begin{enumerate}
	\item describe symmetry groups $\GS$ other than the extended Lorentz group \cite{Butter:2021dtu}. This is essential to treat manifolds with reduced holonomy groups, as they are common in supersymmetric compactifications. Additionally, larger symmetries that go beyond the extended Lorentz group have been considered \cite{Butter:2022iza} to address the problem of the underconstrained Levi-Civita connection.
	\item treat other duality groups, especially the exceptional groups \cite{Hassler:2023axp}. Motivation for this extension is to better understand M-theory and its U-dualities. Moreover, it reveals some central property of the of curvatures in extended geometries. Due to their gauge-for-gauge symmetries, a hierarchy of connections and their curvatures arises. Already at the level of O($d$,$d$) this hierarchy can be anticipated but for the exceptional groups it becomes clear that it is related to the tensor hierarchy of gauged supergravities \cite{deWit:2008ta}.
	\item understand it as a natural lift of Cartan geometry to generalized geometry \cite{Hassler:2024hgq} where it governs the target-space and the world-sheet dynamics of the underlying (gauged) $\sigma$-models.
\end{enumerate}

Following the intuition of Cartan geometry, the starting point is an extended version of the physical space-time $M$ with a flat generalized connection\footnote{A flat connection has only torsion but no curvature.}. We will call it mega-space to hint that it is even larger than the usual extended space used in double/extended field theories. A main insight of \cite{Polacek:2013nla} is that after choosing an appropriate generalized frame on the mega-space, its generalized torsion contains all covariant torsion and curvature tensors of $M$. At this point there arise several questions on the technical implementation. Most important among them are: How to choose the mega-space frame? How to define its generalized torsion? How to extract the relevant curvature components? In previous works these questions have been answered assuming that $\GS$ is an isotropic subgroup of the mega-space's duality group. The present article is concerned with removing the isotropy condition and shows that in this way covariant curvature tensors for the heterotic and type I string arise. Besides the metric, dilaton and the two-form $B$ field from the bosonic string or the NS/NS sector of type II strings, they also contain a non-Abelian gauge potential.

Mathematicians have very similar ideas, clearly predating \cite{Polacek:2013nla}, under the name of symplectic reduction of Courant algebroids \cite{Severa:2017oew,Bursztyn:2005vwa}. In particular \v{S}evera described in his 4\textsuperscript{th} letter to Alan Weinstein \cite{Severa:2017oew} the reduction of exact to transitive Courant algebroids which is the same setting we are interested here. Based on these results, the reduction of curvature tensors has been studies in \cite{Jurco:2015bfs}. Therefore, one might want to conclude that the \PS{} construction is just the index-full version of this reduction. While in some sense this might be true, there is still structure from the \PS{} construction which is not immediately visible in the mathematical literature. Most prominently, is the appearance of additional connections together with their torsions and curvatures from the tensor hierarchy. At the leading, two-derivative, action they will not be relevant but they are very likely central in obtaining higher-derivative corrections. In particular because these corrections are known to be tightly linked to the additional gauge potential appearing for the heterotic string \cite{Bedoya:2014pma,Coimbra:2014qaa}.

We take this as motivation to revisit the existing results about the \PS{} construction and revisit them for non-isotropic symmetry groups $\GS$ to further explore the relation with the heterotic string. As previous experience showed, it is helpful to first understand the underlying algebraic structure which is composed of the symmetry (or equally generalized structure) group $\GS$, the mega-space duality group $\GD$ and the group $\GPS$ furnished by the mega-space frame. Everything else  follows in principle directly from the algebra -- while there is still a bit of work required to see how. Hence, after a quick review of the salient feature of the generalized geometry/double field theory for heterotic strings in section~\ref{sec:reviewhet}, we begin our exploration with the algebraic structure in section~\ref{sec:algebra}. Following this path, we present a more algebraic approach to generalized diffeomorphisms in section~\ref{sec:megaGenDiff}. Compared to the usual way of writing the generalized Lie derivative in terms of the four index $Y$- or $Z$-tensor, it employs an index free notation which was originally introduces for $E_{9(9)}$ exceptional field theory in \cite{Bossard:2017aae}. After these basics are settled, we systematically construct the mega-space frame and discuss its torsion components in section~\ref{sec:genmega-frame}. Until this point, we keep things as general as possible and only in section~\ref{eqn:PSaction} an explicit parameterization for the frame is chosen. It eventually result in the hierarchy of curvature we are after and allows to construct the invariant, low-energy effective action of the heterotic string.

Beyond the excepted importance for higher-derivative corrections, an immediate application for our method is the construction of new heterotic string vacua by employing ideas from generalized dualities and consistent truncations. To demonstrate how this works, we deform the background corresponding to the axial- and vectorial-U(1)-gauging of the SU(2) Wess-Zumino-Witten model in section~\ref{sec:cigarToTrumpet}. The origins of our deformation have been extensively studied as a realization \cite{Bardakci:1990lbc} of the parafermionic CFT \cite{Fateev:1985mm} and their integrable deformations driven by the parafermionic bilinear. Moreover, they can be Wick-rotated to a two-dimensional black-hole \cite{Witten:1991yr}, also called the ``cigar'' geometry and its T-dual ``trumpet'' geometry \cite{Giveon:1994fu}. Finally we conclude in section~\ref{sec:conclusions}.

\section{Review of the heterotic double field theory}\label{sec:reviewhet}
To capture the leading order, low-energy effective action of the heterotic or type I string, double field theory or generalized geometry have proven to be very useful. We will focus on the NS/NS sector with a duality $\GD=\mathrm{O}(d+p,d+q)$. Only the particular choice $p=0$ and $q=496$ are relevant for the ten dimensional effective action of strings. However other cases also appear in the context of reductions and the closely related gauged double field theory \cite{Grana:2012rr}. Therefore, we keep $p$ and $q$ here general. Besides diffeomorphisms and two-form gauge transformations, this duality group also captures a non-abelian gauge symmetry with the gauge group $\GH$. For heterotic/type I strings this gauge group is either E$_8 \times$E$_8$ or SO(32). However, if we consider the first order $\alpha'$ corrections by following \cite{Bedoya:2014pma}, the gauge group additionally contains the SO(1,9) that corresponds to the local Lorentz transformations. Hence, here we keep it general.

We will work with the flux formulation, where all physical degrees of freedom are contained in the generalized frame $E_{\Ac}{}^{\Ic}$ and the generalized dilaton $\Phi$. In addition to being an element of the duality group, the former relates flat indices like $\Ac$ and curved indices like $\Ic$. We decompose them according to
\begin{equation}
	V_{\Ac} = \begin{pmatrix} V_A & V^{\alphai} \end{pmatrix}\,,
	\qquad \text{and} \qquad
	V^{\Ic} = \begin{pmatrix} V^I \\ V_{\mui} \end{pmatrix}\,,
\end{equation}
where the indices $A$, and $I$ describe the fundamental of the subgroup O($d$,$d$), while $\alphan$ and $\mun$ enumerate the basis element of the gauge group $\GH$. To lower/raise these indices, we use the $\eta$-metric
\begin{equation}\label{eqn:heta}
	\heta_{\Ac\Bc} = \begin{pmatrix}
		\eta_{AB} & 0                      \\
		0         & \kappa^{\alphai\betai}
	\end{pmatrix}
\end{equation}
and its inverse. It contains its O($d$,$d$) counter part $\eta_{AB}$ and the inverse of a non-degenerate, invariant metric $\kappa^{\alphai\betai}$ for the gauge group. There are two kinds of local symmetries which act on the frame $E_{\Ac}{}^{\Ic}$. From the left, we have double Lorentz transformations
\begin{equation}
	E'_{\Ac}{}^{\Ic} = \Lambda_{\Ac}{}^{\Bc} E_{\Bc}{}^{\Ic}
\end{equation}
where the matrix $\Lambda_{\Ac}{}^{\Bc}$ is an element of the subgroup O($d+p$)$\times$O($d+q$)$\subset$O($d+p$,$d+q$). This subgroup is fixed by the generalized metric $\cH_{\Ac\Bc} = \cH_{\Bc\Ac}$, resulting in
\begin{equation}
	\Lambda_{\Ac}{}^{\Cc} \Lambda_{\Bc}{}^{\Dc} \cH_{\Cc\Dc} = \cH_{\Ac\Bc}\,.
\end{equation}
As a consequence, the physical degrees for freedom in the frame are contained in the coset O($d+p$,$d+q$)/(O($d+p$)$\times$O($d+q$)). To make contact with the mass-less bosonic field arising from the heterotic string, another branching to GL($d$) reveals that the $(d+p)(d+q)$ degrees of freedom in this coset contain the metric $g_{ij}$, the $B$-field $B_{ij}$, a gauge field $A^{\alphai}_i$ and $p q$ scalars. Note that for $p=0$ the scalars vanish and one recovers the expected field content. In this case, the generalized metric reads\cite{Hohm:2011ex}
\begin{equation}\label{eqn:cHparam}
	\cH^{\Ic\Kc} = E_{\Ac}{}^{\Ic} \cH^{\Ac\Bc} E_{\Bc}{}^{\Kc} = \begin{pmatrix}
		g^{jk}                & - g^{jl} c_{lk}                                                    & - g^{jl} A_l^{\betai}                                      \\
		-g^{kl} c_{lj} \quad  & g_{jk} + c_{lj} g^{lm} c_{mk} + A_j{}^{\gammai} A_{k\gammai} \quad & c_{lj} g^{lm} A_m^{\betai} + A_j^{\betai}                  \\
		-g^{kl} A_l^{\alphai} & c_{lk} g^{lm} A_m^{\alphai} + A_k^{\alphai}                        & \kappa^{\alphai\betai} + A_l^{\alphai} g^{lm} A_m^{\betai}\end{pmatrix}
\end{equation}
with
\begin{equation}
	c_{jk} = B_{jk} + \tfrac12 A_j^{\alphai} A_{k\alphai}\,.
\end{equation}

Generalized diffeomorphisms act from the right on the frame through the generalized Lie derivative
\begin{equation}
	\genLieHet_\xi E_{\Ac}{}^{\Kc} = \xi^{\Ic} \partial_{\Ic} E_{\Ac}{}^{\Kc} + \left( \partial^{\Kc} \xi_{\Ic} - \partial_{\Ic} \xi^{\Kc} \right) E_{\Ac}{}^{\Ic} + \xi^{\gammai} E_{\Ac}{}^{\alphai} f_{\gammai\alphai}{}^{\betai} \delta_{\betai}^{\Kc}\,.
\end{equation}
The last term in this expression is a twist, or torsion, that contains the structure coefficients of the gauge group $\GH$. Finally, there is the generalized dilaton $\Phi$ defined by
\begin{equation}
	\Phi = \phi - \tfrac12 \log \sqrt{g}
\end{equation}
in terms of the dilaton $\Phi$ and the determinate of the metric $g$. It usually appears in form of the exponent $e^{-2\Phi}$ whose generalized Lie derivative is given by
\begin{equation}
	\genLieHet_\xi e^{-2\Phi} = \xi^{\Ic} \partial_{\Ic} e^{-2\Phi} + \partial_{\Ic} \xi^{\Ic} e^{-2 \Phi}\,.
\end{equation}
After all physical fields are collected, we are ready to present their two-derivative low-energy effective action\cite{Geissbuhler:2013uka}
\begin{equation}\label{eqn:ShetDFT}
	\begin{aligned}
		S = V \int \dd^d x \, e^{-2\Phi} \Bigl( & 2 \cH^{\Ac\Bc} D_{\Ac} F_{\Bc} - \cH^{\Ac\Bc} F_{\Ac} F_{\Bc} + \tfrac14 \cH^{\Ac\Dc} F_{\Ac\Bc\Cc} F_{\Dc}{}^{\Bc\Cc}                                                              \\
		                                        & - \tfrac1{12} \cH^{\Ac\Bc} \cH^{\Cc\Dc} \cH^{\Ec\Fc} F_{\Ac\Cc\Ec} F_{\Bc\Dc\Fc} \underline{- \tfrac16 F^{\Ac\Bc\Cc} F_{\Ac\Bc\Cc} + F^{\Ac} F_{\Ac} - 2 D_{\Ac} F^{\Ac}} \Bigr)\,,
	\end{aligned}
\end{equation}
which is written in terms of the generalized fluxes
\begin{align}\label{eqn:FAcBcCc}
	F_{\Ac\Bc\Cc} & = \genLieHet_{E_{\Ac}} E_{\Bc}{}^{\Kc} E_{\Cc\Kc} = 3 D_{[\Ac} E_{\Bc}{}^{\Ic} E_{\Cc]\Ic} + E_{\Ac\alphai} E_{\Bc\betai} E_{\Cc\gammai} f^{\alphai\betai\gammai}\,, \\
	F_{\Ac}       & = \genLieHet_{E_{\Ac}} e^{2 \Phi} e^{-2 \Phi} = 2 D_{\Ac} \Phi - \partial_{\Ic} E_{\Ac}{}^{\Ic}\,.
\end{align}
Note that out of convenience, we have introduced the flat derivative
\begin{equation}
	D_{\Ac} = E_{\Ac}{}^{\Ic} \partial_{\Ic}\,.
\end{equation}
Furthermore, the underlined terms in \eqref{eqn:ShetDFT} are annihilated by the section condition
\begin{equation}\label{eqn:hetSC}
	\partial_{\Ic} \, \cdot \, \partial^{\Ic} \, \cdot \, = 0\,.
\end{equation}
It is usually imposed on arbitrary combinations of fields and parameters for gauge transformations. However, in section~\ref{sec:cigarToTrumpet} we will see that it can be relaxed under specific circumstances without scarifying the consistency of the theory.

\section{Algebra}\label{sec:algebra}
We already explained that the \PS{} construction utilizes the mega-space to geometrize generalized diffeomorphisms and local gauge transformations on the physical space $M$. To make this idea more explicit, one first has to understand how the respective symmetries are related. We assume that $M$ is $d$-dimensional and its generalized diffeomorphisms furnish the duality group $\GD=\Odd{}$. There are also local transformations that generate the structure group $\GS$. An obvious choice for $\GS$ would be the double Lorentz group but we want to keep it general. Its adjoint representation is embedded into $n\times n$ matrices, where $n=\dim\GS$. This clearly always works because $\GS \subset \GL[n]$. The main idea behind the \PS{} construction is that $\GD$ and $\GS$ arise from generalized diffeomorphisms on the $(d+n)$-dimensional mega-space, governed by the group $\GM=\Odd[d+n]$. To see how these three groups are related, consider the branching
\begin{equation}\label{eqn:initialBranching}
	\Odd[d+n] \rightarrow \Odd[d] \times \GL[n] \,.
\end{equation}
The left-hand side is generated by
\begin{equation}\label{eqn:levelDecompGenList}
	\Rt^{\alpha\beta},\, \Rt^\alpha_A,\, K_{AB},\, K_\alpha^\beta,\, R_\alpha^A, \quad \text{and} \quad R_{\alpha\beta}
\end{equation}
with indices $A=1,\ldots,2 d$ and $\alpha=1,\ldots,n$. They of course include the generators of the right-hand side, $K_{AB}$ and $K_\alpha^\beta$, governed by the non-trivial commutators
\begin{align}\label{eqn:commKABKCD}
	[ K_{AB}, K_{CD} ]                  & = - \tfrac12 ( \eta_{AC} K_{BD} - \eta_{AD} K_{BC} + \eta_{BD} K_{AC} - \eta_{BC} K_{AD} ) = 2 \eta_{[A|[C} K_{D]|B]}\,, \\
	[ K_\alpha^\beta, K_\gamma^\delta ] & = \delta_\alpha^\delta K_\gamma^\beta - \delta_\gamma^\beta K_\alpha^\delta\,,
\end{align}
where $\eta_{AB}$ denotes the invariant metric of $\Odd[d]$. To obtain the commutators of the remaining generators in \eqref{eqn:levelDecompGenList}, we first assign a grading, called level $\ell$, to each generator according to the position of its indices. More specifically, each lower/raised Greek index contributes with $-1$/$+1$. Equipped with this convention, we find that the $K$-generators furnish level zero. All levels beyond $|\ell|>1$ are freely generated, namely
\begin{align}
	[ \Rt^\alpha_A , \Rt^\beta_B ] & = \eta_{AB} \Rt^{\alpha\beta}\,, \qquad\text{and}              \\\label{eqn:commR1R1}
	[ R^A_\alpha, R^B_\beta ]      & = \eta^{AB} R_{\alpha\beta} - 2 K^{AB} \kappa_{\alpha\beta}\,.
\end{align}

We will see that the last term in the second line is crucial to make contact to the heterotic string from the original \PS{} construction presented in \cite{Polacek:2013nla,Butter:2022iza,Hassler:2023axp}. Here, we take $\kappa_{\alpha\beta}$ be symmetric. Although it is not a generator, according to its index position it still has grading $-2$. The adjoint action of the $\Odd[d]$-generator $K_{AB}$ reads
\begin{align}
	[ \Rt^\alpha_A, K_{BC} ] & = \phantom{-} (K_{BC})_A{}^D \Rt^\alpha_D\,, \qquad \text{and} \\\label{eqn:commR1K}
	[ K_{AB}, R^C_\gamma ]   & = - (K_{AB})^C{}_D R^D_\gamma\,,
\end{align}
where the indices $A$, $B$, $\dots$ are raised/lowered with $\eta_{AB}$ and
\begin{equation}
	(K_{AB})_{CD} = \eta_{[A|C} \eta_{D|B]}
\end{equation}
mediates the fundamental action of $\Odd[d]$. It is fully fixed by the generators' index structure. By combining \eqref{eqn:commR1R1} and \eqref{eqn:commR1K}, we find the remaining non-trivial commutators
\begin{align}\label{eqn:commR1R2}
	[ R^A_\alpha, R_{\beta\gamma} ]       & = - 2 \kappa_{\alpha[\beta} R_{\gamma]}^A\,, \qquad \text{and} \\\label{eqn:commR2R2}
	[ R_{\alpha\beta}, R_{\gamma\delta} ] & = - 4 \kappa_{[\alpha| [\gamma} R_{\delta]|\beta]}
\end{align}
of all negative level generators. At this point we see that
\begin{equation}\label{eqn:KABandRs}
	\gPS = \{ K_{AB}\,, R^A_\alpha\,, R_{\alpha\beta} \}
\end{equation}
generates a subgroup of $\GM$ that we will denote as $\GPS$ because of its central role in the construction of the generalized frame on the mega-space as outlined in section~\ref{sec:genmega-frame}. By the Jacobi identity, the relation between positive and negative level generators is fixed to
\begin{equation}
	[ \Rt^\alpha_A , R^B_\beta ] = - \delta_A^B \left( K^\alpha_\beta -\tfrac12 \Rt^{\alpha\gamma} \kappa_{\gamma\beta} \right) - 2 \delta^\alpha_\beta K_A{}^B \,.
\end{equation}
In the same vein, all other non-vanishing commutators arise. They are summarized in appendix~\ref{app:commutators}.

\subsection{Highest weight representations}\label{sec:vermamodule}
After the algebra that underlies the \PS{} construction is fixed, we will focus on its representations. In particular, we are interested in the Verma modules of the fundamental representation, which is characterized by the highest weight state $|^\alpha\rangle$. Accordingly, $|^\alpha\rangle$ is annihilated by the raising operators $\Rt^\alpha_A$ and $\Rt^{\alpha\beta}$, while the lowering operators $R^A_\alpha$ and $R_{\alpha\beta}$ generate the full representation. For the dual state $\langle_\alpha|$, with
\begin{equation}
	\langle_\alpha|^\beta \rangle = \delta_\alpha^\beta\,,
\end{equation}
the situation is reversed as it is annihilated by the lowering operators. Both states do not carry any $\Odd[d]$ indices and are therefore annihilated by $K_{AB}$, too. Consequentially, the last thing to fix is the action of $K_\alpha^\beta$. It is given by
\begin{equation}
	K_\alpha^\beta |^\gamma\rangle = \delta_\alpha^\gamma |^\beta\rangle\,.
\end{equation}

These rules are sufficient to evaluate any ``correlator'' of the form $\langle_\alpha | \dots |^\beta\rangle$, where $\dots$ denotes some arbitrary combination of generators. Like in quantum mechanics or quantum field theory, one only has to bring the operators $\dots$ into normal order. Using this technique, we define the normalized states
\begin{equation}
	\begin{aligned}
		|^A \rangle      & = \phantom{-}\tfrac{1}{n} R_\alpha^A |^\alpha \rangle\,,             &
		\langle_A |      & = -\tfrac{1}{n} \langle_\alpha | \Rt^\alpha_A\,,                       \\
		|_\alpha \rangle & = -\tfrac{1}{n-1} R_{\alpha\beta} |^\beta\rangle\,, \qquad\text{and} &
		\langle^\alpha | & = -\tfrac{1}{n-1} \langle_\beta | \Rt^{\beta \alpha} \,.
	\end{aligned}
\end{equation}
They form the level decomposition of the fundamental representation and it is therefore convenient to combine them into
\begin{equation}\label{eqn:decompAh}
	|^{\Ah}\rangle = \begin{pmatrix} |^\alpha\rangle & |^A\rangle & |_\alpha\rangle \end{pmatrix} \qquad \text{and} \qquad
	\langle_{\Ah}| = \begin{pmatrix} \langle_\alpha| & \langle_A | & \langle^\alpha | \end{pmatrix}{}^T,
\end{equation}
which are dual, namely satisfy
\begin{equation}
	\langle_{\Ah}|^{\Bh} \rangle = \delta_{\Ah}^{\Bh}\,,
\end{equation}
due to the definition of the states above. This eventually allows us to compute matrix representations for all generators introduced in the last subsection. In particular, we find
\begin{align} \label{eqn:matrixKs}
	\langle_{\Ah} | K_\gamma^\delta |^{\Bh} \rangle    & = \begin{pmatrix}
		                                                       \delta_\alpha^\delta \delta_\gamma^\beta & 0 & \kappa_{\gamma(\alpha} \delta_{\beta)}^\delta \\
		                                                       0                                        & 0 & 0                                             \\
		                                                       0                                        & 0 & -\delta_\gamma^\alpha \delta_\beta^\delta
	                                                       \end{pmatrix}\,,                         &
	\langle_{\Ah} | K_{CD} |^{\Bh} \rangle             & = \begin{pmatrix}
		                                                       0 & 0                        & 0 \\
		                                                       0 & \eta_{A[C} \delta_{D]}^B & 0 \\
		                                                       0 & 0                        & 0
	                                                       \end{pmatrix}\,,                                                                                        \\
	\langle_{\Ah} | \Rt^\gamma_C |^{\Bh} \rangle       & = \begin{pmatrix}
		                                                       0 & -\delta_\alpha^\gamma \delta_C^B & 0                             \\
		                                                       0 & 0                                & \delta_\beta^\gamma \eta_{AC} \\
		                                                       0 & 0                                & 0
	                                                       \end{pmatrix}\,,                                                 &
	\langle_{\Ah} | R_\gamma^C |^{\Bh} \rangle         & = \begin{pmatrix}
		                                                       0                              & 0                                & 0                               \\
		                                                       \delta_\gamma^\beta \delta_A^C & 0                                & \kappa_{\beta\gamma} \delta_A^C \\
		                                                       0                              & - \delta_\gamma^\alpha \eta^{BC} & 0
	                                                       \end{pmatrix}\,,                     \\ \label{eqn:matrixR2s}
	\langle_{\Ah} | \Rt^{\gamma\delta} |^{\Bh} \rangle & = \begin{pmatrix}
		                                                       0 & 0 & -2 \delta_{[\alpha}^\gamma \delta_{\beta]}^\delta \\
		                                                       0 & 0 & 0                                                 \\
		                                                       0 & 0 & 0
	                                                       \end{pmatrix}\,,\qquad\text{and}                                                            &
	\langle_{\Ah} | R_{\gamma\delta} |^{\Bh} \rangle   & = \begin{pmatrix}
		                                                       0                                                 & 0 & 0                                               \\
		                                                       0                                                 & 0 & 0                                               \\
		                                                       -2 \delta_{[\gamma}^\alpha \delta_{\delta]}^\beta & 0 & - \delta_{[\gamma}^\alpha \kappa_{\delta]\beta}\end{pmatrix}\,.
\end{align}
One can easily check that all these generators leave the $\Odd[d+n]$ metric
\begin{equation}
	\eta_{\Ah\Bh} = \begin{pmatrix}
		- \kappa_{\alpha\beta} & 0         & \delta_\alpha^\beta \\
		0                      & \eta_{AB} & 0                   \\
		\delta_\beta^\alpha    & 0         & 0
	\end{pmatrix}
\end{equation}
invariant. Note that because we constructed the matrices in \eqref{eqn:matrixKs}-\eqref{eqn:matrixR2s} starting from a highest weight state and its dual, the first row of all lowering operators is zero.

\subsection{Heterotic basis}\label{sec:hetBasis}
Until now, we have not put any additional constraints on $\kappa_{\alpha\beta}$ despite it being symmetric with respect to its two indices. By a suitable choice of basis vectors, it is always possible to bring it into the form
\begin{equation}
	\kappa_{\alpha\beta} = \begin{pmatrix}
		\kappa_{\alphai\betai} & 0 \\
		0                      & 0\end{pmatrix}
\end{equation}
after decomposing the indices corresponding to the structure group $\alpha = \begin{pmatrix} \alphai & \alphan \end{pmatrix}$. In this adapted basis, $\kappa_{\alphai\betai}$ is non-degenerate and we assume it has signature $(p,q)$. Therefore, its indices like $\alphai$ run from 1 to $p+q$, the rank of $\kappa_{\alpha\beta}$, while $\alphan$ continues from $p+q+1$ to $n$. Therefore, a natural decomposition of $n$ is
\begin{equation}
	n = p+ q + r\,.
\end{equation}
After this refinement, there is an interesting, alternative choice of basis vectors
\begin{equation}\label{eqn:hetStates}
	\begin{aligned}
		|^{\alphai} \hrangle & = |^{\alphai}\rangle - \kappa^{\alphai\betai} |_{\betai} \rangle \,, &
		\hlangle_{\alphai} | & = \langle_{\alphai}|\,,                                                                                                                                         \\
		|^{\alphan} \hrangle & = |^{\alphan}\rangle\,,                                              &
		\hlangle_{\alphan} | & = \langle_{\alphan}|\,,                                                                                                                                         \\
		|^A\hrangle          & = |^A\rangle\,,                                                      & \hlangle_A|         & = \langle_A|\,,                                                    \\
		|_{\alphai} \hrangle & = |_{\alphai}\rangle\,,                                              & \hlangle^{\alphai}| & = \langle^{\alphai}| + \kappa^{\alphai\betai} \langle_{\betai}|\,,
		\\
		|_{\alphan}\hrangle  & = |_{\alphan}\rangle\,,\qquad\text{and}\qquad                        & \hlangle^{\alphan}| & = \langle^{\alphan}|\,.
	\end{aligned}
\end{equation}
for the fundamental representation and its dual. As we will see in the following, it puts the emphasis on a $\GH=\mathrm{O}(d+p,d+q)$, subgroup of mega-space generalized diffeomorphisms. For $p=0$, the latter appears in heterotic double field theory and thereby justify the name ``heterotic basis''. To see how this works, first note that $K_{AB}$ keeps its matrix form
\begin{equation}
	\hlangle_{\Ah}| K_{CD} |^{\Bh}\hrangle =  \langle_{\Ah}| K_{CD} |^{\Bh}\rangle\,
\end{equation}
after the change of basis. In particular, its action is only non-trivial on $|^A\hrangle$ which we combine with $|^{\alphai}\hrangle$ to
\begin{align}
	|^{\Ac} \hrangle & = \begin{pmatrix} |^A\hrangle\quad & |_{\alphai}\hrangle
	                     \end{pmatrix}\,.
	\intertext{This new state, with the dual}
	\hlangle_{\cA} | & = \begin{pmatrix} \hlangle_A|\quad & \hlangle^{\alphai}| \end{pmatrix}{}^T\,,
\end{align}
describes the fundamental representation of $\GH$. Its generators are given by
\begin{equation}
	\Khet_{\Ac\Bc} = \begin{pmatrix}
		\phantom{-}K_{AB}        \quad & \tfrac12 R_A^{\betai}        \\
		-\tfrac12 R^{\alphai}_B \quad  & - \tfrac12 R^{\alphai\betai}\end{pmatrix}\,,
\end{equation}
where we use the inverse of $\kappa^{\alphai\betai}$ of $\kappa_{\alphai\betai}$ to raise the respective indices. With this definition, one find the commutators
\begin{equation}\label{eqn:commKhetKhet}
	\boxed{%
	[ \Khet_{\Ac\Bc}, \Khet_{\Cc\Dc} ] = 2 \eta_{[\Ac|[\Cc} \Khet_{\Dc]|\Bc]}}
\end{equation}
from the previous relations with the heterotic $\eta$-matrix defined in \eqref{eqn:heta}.

At this point, we conclude that the newly introduced $\Khet_{\Ac\Bc}$ generate the group O$(d+p,d+q)$. For the remaining generators of $\GPS$, we combine
\begin{equation}\label{eqn:hetRs}
	\Rhet^{\Bc}_{\alphan} = \begin{pmatrix} R^B_{\alphan}\,\, & R_{\alphan\betai} \end{pmatrix} \qquad \text{while keeping} \qquad
	\Rhet_{\alphan\betan} = R_{\alphan\betan}
\end{equation}
as it is. Remarkably, this choice show that $\GPS$ still preserves the $\mathbb{Z}_3$ grading which is a hallmark of the untwisted \PS{} construction. Therefore, in addition to \eqref{eqn:commKhetKhet}, we only have the two non-trivial commutators
\begin{equation}
	\boxed{%
		\begin{aligned}
			\relax[\Khet_{\Ac\Bc}, \Rhet^{\Cc}_{\gamman} ]  & = - \delta_{[\Ac}^{\Cc} \Rhet_{\Bc]\gamman}\,, \qquad \text{and} \\
			[ \Rhet^{\Ac}_{\alphan}, \Rhet^{\Bc}_{\betan} ] & = \heta^{\Ac\Bc} \Rhet_{\alphan\betan}\,.
		\end{aligned}}
\end{equation}
Note that $\heta^{\Ac\Bc}$ is the inverse of $\heta_{\Ac\Bc}$ and that both are used to raise and lower capital, calligraphic (heterotic) indices. In analogy with \eqref{eqn:hetRs}, we define the tilded generators
\begin{equation}
	\Rthet^{\alphan}_{\Bc} = \begin{pmatrix} \Rt^{\alphan}_B \quad & \tfrac12 \Rt^{\alphan\betai} - K^{\betai\alphan} \end{pmatrix}\,, \qquad \text{and} \qquad \Rthet^{\alphan\betan} = \Rt^{\alphan\betan}\,.
\end{equation}
As one might expect, they again fulfill the untwisted commutation relations, most importantly
\begin{equation}
	[ \Rthet_{\Ac}^{\alphan}, \Rhet^{\Bc}_{\betan} ] = -\delta^{\Bc}_{\Ac} \Khet_{\betan}^{\alphan} - 2 \delta_{\betan}^{\alphan} \Khet_{\Ac}{}^{\Bc}
	\qquad \text{with} \qquad
	\Khet_{\betan}^{\alphan} = K_{\betan}^{\alphan}\,.
\end{equation}
Together with the generators of $\GPS$, they form the Lie algebra
\begin{equation}
	\gM' = \left\{ \Rthet^{\alphan\betan},\, \Rthet^{\alphan}_{\Ac},\, \Khet_{\alphan}^{\betan},\, \Khet_{\Ac\Bc},\, \Rhet^{\Ac}_{\alphan},\, \Rhet_{\alphan\betan} \right\}
\end{equation}
which generates the subgroup $\GM'=\mathrm{O}(d+p+r, d+q+r)$ of $\GM$. The crucial property of this group is that all its elements preserves the states $|^{\alphai} \hrangle$ and its dual, \begin{equation}\label{eqn:lowerblock}
	g |^{\alphai} \hrangle = |^{\alphai} \hrangle \,,
	\qquad \text{and} \qquad
	\hlangle_{\alphai} | g = \hlangle_{\alphai} |  \qquad \forall \, g \in \GM'\,.
\end{equation}

In general the dimension of $\GM'$ is smaller than that of $\GM$. Hence, we have to have a look at the remaining generators. We choose them to be the orthogonal complement -- saying that any trace of them with any element of $\gM'$ vanishes. Working out the respective matrix representations explicitly, we obtain
\begin{align}
	\Rthet^{\alphai\betai}   & = K^{[\alphai\betai]} + \tfrac12 R^{\alphai\betai}\,,         &
	\Khet_{\alphan}^{\betai} & = K_{\alphan}^{\betai} - R_{\alphan}{}^{\betai}\,,            &
	\Khet_{\alphai}^{\betan} & = K_{\alphai}^{\betan} - \tfrac12 \Rt_{\alphai}{}^{\betan}\,,
\end{align}
and
\begin{equation}
	\label{eqn:Rtprimed}
	\Rthet^{\alphai}_{\Bc}    = \begin{pmatrix} \Rt_B^{\alphai} - R^{\alphai}_B \qquad &  \tfrac12 \Rt^{\alphai\betai} - K^{\alphai\betai} - R^{\alphai\betai} \end{pmatrix}
	\qquad \text{with} \qquad K^{\alphai\betai} := K_{\gammai}^{\alphai} \kappa^{\betai\gammai} \,.
\end{equation}
They annihilate $|^{\alphan}\hrangle$, $|^{\Ac}\hrangle$, $|_{\alphan}\hrangle$ and their respective dual states $\hlangle_{\alphan}|$, $\hlangle_{\Ac}|$, $\hlangle^{\alphan}|$. Collecting them into the set
\begin{equation}\label{eqn:fpgenerators}
	\fp = \left\{ \Rthet^{\alphai\betai}\,, \hK_{\alphan}^{\betai}\,, \hK_{\alphai}^{\betan}\,, \hRt^{\alphai}_{\Bc} \right\}\,,
\end{equation}
one can show that $\fp$ forms a representation under the adjoint-action of $\gM'$, namely
\begin{equation}
	[ \gM', \fp ] \subset \fp\,.
\end{equation}
Note that the generators $\Rthet^{\alphai\betai} \in \fp$ form the Lie algebra of O$(p,q)$. To summarize our efforts until this point: We discovered that starting from the non-vanishing $\kappa_{\alpha\beta}$ the original branching \eqref{eqn:initialBranching} of $\GM$ gets refined to
\begin{equation}
	\mathrm{O}(d+n,d+n) \rightarrow \mathrm{O}(d+p+r, d+q+r) \rightarrow \mathrm{O}(d+p,d+q) \times \mathrm{GL}(r)\,.
\end{equation}

Later we will see in particular that $\GPS$ plays a central role in capturing the geometry of the physical space. Therefore, it is useful when we have a straightforward way to project from $\gM$ to $\gPS$. To this end, we use the states
\begin{equation}
	|_{\Ad} \hrangle = \begin{pmatrix} |_A \hrangle \quad & |^\alpha\hrangle \end{pmatrix} = \begin{pmatrix} |_{\Ac}\hrangle \quad & |^{\alphan} \hrangle \end{pmatrix} \,,
	\qquad \text{and} \qquad
	\hlangle_{\Ad}| = \begin{pmatrix} \hlangle_A | \quad & \hlangle^\alpha | \end{pmatrix}{}^T = \begin{pmatrix} \hlangle_{\Ac} | \quad & \hlangle^{\alphan} | \end{pmatrix}{}^T\,.
\end{equation}

\subsection{Generalized structure group and generalized metric}\label{sec:GS}
A central object of the \PS{} construction is the generalized structure group $\GS$. Until now, we have just stated that it is embedded into $\GL[n]$, where $n=\dim\GS$. Now, it is time to make this statement more precise.

Assume the Lie algebra of $\GS$ is spanned by the generators $\hatt_\alpha$ with the commutators
\begin{equation}\label{eqn:commtalphabeta}
	[ \hatt_\alpha, \hatt_\beta ] = - f_{\alpha\beta}{}^\gamma \hatt_\gamma\,.
\end{equation}
The embedding of these generators into the Lie algebra of $\GL[n]$ is given by
\begin{equation}
	K_\alpha = f_{\alpha\beta}{}^\gamma K_\gamma^\beta\,.
\end{equation}
Furthermore, we require that the action of this group preserves the subgroup $\GPS$ generated by \eqref{eqn:KABandRs}. This is the case only if
\begin{equation}\label{eqn:kappainvariant}
	f_{\alpha(\beta}{}^\delta \kappa_{\gamma)\delta} = 0
\end{equation}
holds, or equally $\kappa_{\alpha\beta}$ has to be invariant under the action of $\GS$. In this case, the relevant commutators simplify to
\begin{align}\label{eqn:genIPRthetii}
	[ K_\alpha, K_\beta ]         & = - f_{\alpha\beta}{}^\gamma K_\gamma\,,                    \\
	[ K_\alpha, R^B_\beta ]       & = - f_{\alpha\beta}{}^\gamma R^B_\gamma\,, \qquad\text{and} \\
	[ K_\alpha, R_{\beta\gamma} ] & = 2 f_{\alpha[\beta}{}^\delta R_{\gamma]\delta}\,.
\end{align}
After decomposing the Greek indices of the structure coefficients, we find that \eqref{eqn:kappainvariant} implies
\begin{equation}\label{eqn:zerofs}
	f_{\alpha\betan}{}^{\gammai} = 0\,.
\end{equation}
In this case, there are four free contributions to the structure coefficients, namely
\begin{equation}
	f_{\alphai\betai}{}^{\gammai}\,, \quad f_{\alphai\betai}{}^{\gamman}\,, \quad f_{\alphai\betan}{}^{\gamman}\,, \quad \text{and} \quad f_{\alphan\betan}{}^{\gamman}\,.
\end{equation}
Looking at the Jacobi identity $3 f_{[\alpha\beta}{}^\epsilon f_{\gamma]\epsilon}{}^\delta = 0$ for \eqref{eqn:commtalphabeta}, we find that $f_{\alphai\betai}{}^{\gammai}$ and $f_{\alphan\betan}{}^{\gamman}$ have to satisfy respective Jacobi identities on their own due to \eqref{eqn:zerofs}.

Instead of just identifying $\hatt_\alpha$ and $K_\alpha$, we want to have the most general embedding with a non-trivial action on the subgroup generated by $\gPS$. This is achieved by decomposing
\begin{equation}\label{eqn:deftalpha}
	\hatt_\alpha = K_\alpha + t_\alpha \qquad \text{with} \qquad \gPS \ni t_\alpha =
	(t_\alpha)_{\Bc\Cc} \Khet^{\Bc\Cc} + (t_\alpha)_{\Bc}{}^{\gamman} \Rhet^{\Bc}_{\gamman} + \tfrac12 (t_\alpha){}^{\betan\gamman} \Rhet_{\gamman\betan}\,.
\end{equation}
The constants $(t_\alpha)_{\Bc\Cc}$, $(t_\alpha)_{\Bc}{}^{\gamman}$ and $(t_\alpha)^{\betan\gamman}$ cannot be chosen freely because \eqref{eqn:commtalphabeta} additionally implies
\begin{align}\label{eqn:representation}
	2 (t_{[\alpha})_{\Cc}{}^{\Ec} (t_{\beta]})_{\Ec}{}^{\Dc}                                                                                                                                            & = - f_{\alpha\beta}{}^{\gamma} (t_{\gamma})_{\Cc}{}^{\Dc}\,,                       \\
	2 (t_{[\alpha})_{\Cc}{}^{\epsilonn} f_{\beta]\epsilonn}{}^{\deltan} - 2 (t_{[\alphan})_{\Ec}{}^{\deltan} (t_{\beta]})_{\Cc}{}^{\Ec} + f_{\alpha\beta}{}^{\epsilon} (t_{\epsilon})_{\Cc}{}^{\deltan} & = 0 \,,                                                          \label{eqn:tJac1} \\
	4 (t_{[\alpha})^{\epsilonn[\deltan} f_{\beta]\epsilonn}{}^{\gamman]} + f_{\alpha\beta}{}^{\epsilon} (t_{\epsilon})^{\gamman\deltan}                                                                 & = 2 (t_{[\alpha})^{\Ec\gamman} (t_{\beta]})_{\Ec}{}^{\deltan}\,.\label{eqn:tJac2}
\end{align}
The first, \eqref{eqn:representation}, requires that $(t_{\alpha})_{\Bc}{}^{\Cc}$ is a representation of $\GS$ on the subspace spanned by $|_{\Ac}\hrangle$. This allows us to introduce an exterior derivative action on the elements of $\gPS$ by
\begin{equation}
	\dd \, \cdot \, = \theta^\alpha [ K_\alpha + (t_\alpha)_{\Bc\Cc} \Khet^{\Bc\Cc}, \, \cdot \, ] \,.
\end{equation}
Its nilpotency requires $\dd \theta^\alpha = - \tfrac12 f_{\beta\gamma}{}^\alpha \theta^\beta\wedge\theta^\gamma$. Finally, we introduce the $\gPS$ valued one forms
\begin{align}
	\tau_1     & = (t_\alpha)_{\Bc}{}^{\gamman} \Rhet^{\Bc}_{\gamman} \, \theta^{\alpha}\,,
	           & \text{and}                                                                     &                   &
	\tau_2     & = \tfrac12 (t_\alpha)^{\betan\gamman} \Rhet_{\gamman\betan} \, \theta^{\alpha}
	\intertext{to rewrite the conditions \eqref{eqn:tJac1}-\eqref{eqn:tJac2} as}
	\dd \tau_1 & = 0 \,,                                                                        & \qquad \text{and} &   &
	\dd \tau_2 & = 2\, \tau_1 \wedge \tau_1\,.
\end{align}
It shows that $\tau_1$ has to be closed and the wedge product $\tau_1 \wedge \tau_1$ has to be exact. A trivial solution for this problem is $\tau_1 = \dd \varphi$ and $\tau_2 = 2 \varphi \wedge \dd\varphi$ for an arbitrary one-form $\varphi$.

We will show now that in this case, we can define a generalized metric on the mega-space which is stabilized by $\GS$. It is defined by the map $\cH: R_1 \rightarrow R_1$ -- from the highest weight representation $R_1$ we constructed in section~\ref{sec:vermamodule} to itself -- that has to satisfy the following properties,
\begin{enumerate}
	\item cubes to itself, $\cH^3 = \cH$,
	\item is symmetric, $\eta_{\Ah\Ch} \cH^{\Ch}{}_{\Bh} = \eta_{\Bh\Ch} \cH^{\Ch}{}_{\Ah}$, and
	\item $\cH(t_\alpha) = 0$\,.
\end{enumerate}
They can be understand as natural extensions of the generalized metric's properties in double field theory or generalized geometry. In particular, they allow to define the two projectors
\begin{equation}
	P^{\Ah}{}_{\Bh} = \frac12\left( \cH^{\Ah}{}_{\Ch} \cH^{\Ch}{}_{\Bh} + \cH^{\Ah}{}_{\Bh}\right)\,, \qquad \text{and} \qquad
	\Pb^{\Ah}{}_{\Bh} = \frac12\left( \cH^{\Ah}{}_{\Ch} \cH^{\Ch}{}_{\Bh} - \cH^{\Ah}{}_{\Bh}\right)\,.
\end{equation}
Moreover, we require that $\cH$ is invariant under the action of $\GS$,
\begin{equation}\label{eqn:genMetricInvar}
	[ \hatt_\alpha, \cH ] = 0\,.
\end{equation}

Implementing the properties for the $\cH$ enumerated above, results in the only non-vanishing contributions
\begin{equation}
	\cH_{\Ad\Bd} = \begin{pmatrix}
		\cH_{\Ac\Bc}                      & \cH_{\Ac\Cc} \varphi^{\Cc\betan} \\
		\cH_{\Bc\Cc} \varphi^{\Cc\alphan} & \cH^{\alphan\betan}
	\end{pmatrix}
	\qquad\text{with}\qquad
	\cH^{\Ac}{}_{\Cc} \cH^{\Cc}{}_{\Bc} = \delta^{\Ac}_{\Bc}\,,
	\quad\text{and}\quad
	\cH^{\alphan\betan} = \varphi^{\Cc\alphan} \varphi^{\Dc\betan} \cH_{\Cc\Dc}\,.
\end{equation}
Here $\cH_{\Ac\Bc}$ is the generalized metric valued in the coset O($d+p$,$d+q$)/O($d+p$)$\times$O($d+q$) and its indices are raised with the inverse $\heta^{\Ac\Bc}$ of $\heta_{\Ac\Bc}$ defined in \eqref{eqn:heta}. For the invariance condition \eqref{eqn:genMetricInvar} to hold, we further need to fix
\begin{equation}
	(t_\alpha)_{\Bc}{}^{\gamman} = - f_{\alpha\deltan}{}^{\gamman} \varphi_{\Bc}{}^{\deltan} - (t_\alpha)_{\Bc}{}^{\Dc} \varphi_{\Dc}{}^{\gamman}
\end{equation}
which remarkably translates directly to
\begin{equation}
	\tau_1 = \dd \varphi\,, \qquad \text{with} \qquad
	\varphi = \varphi_{\Ac}{}^{\betan} \Rhet^{\Ac}_{\betan}\,.
\end{equation}
Note that it is always possible to perform a transformation in $\gPS$ that removes $\varphi$. Therefore, we can work without loss of generality with
\begin{equation}\label{eqn:genMetricStandard}
	\cH_{\Ad\Bd} = \begin{pmatrix}
		\cH_{\Ac\Bc} & 0 \\
		0            & 0
	\end{pmatrix}\,.
\end{equation}

\section{Generalized diffeomorphisms on the mega-space}\label{sec:megaGenDiff}
Eventually, we will use the algebra derived in the last section to compute the generalized Lie derivative $\genLieM$ and therewith define generalized diffeomorphisms on the mega-space. There are different ways to approach this task. Here, we advertise an index-free method, although it might look more difficult in the beginning as it is not the standard route taken in the literature. There are however several advantages that justify this choice: First, as we will see later, it provides more elegant and concise derivations. Second, it can be more easily extended to other duality algebra groups as shown recently in \cite{Hassler:2023axp}. We define the index-free version of the generalized Lie derivative as \cite{Bossard:2017aae}
\begin{equation}\label{eqn:genLie}
	\genLieM_{\langle U|} \langle V| = \langle V |\langle U |\partial V \rangle + \langle V|\langle U| \Zop |\partial U \rangle
\end{equation}
with an appropriately chosen $\Zop$. Moreover, we denote with $\partial V$ the partial derivative acting on $\langle V|$ and adopt the convention
\begin{equation}
	\langle U_1 | \langle U_2 | A \otimes B |V_2 \rangle |V_1 \rangle = \langle U_2 | B | V_2 \rangle \langle U_1 | A | V_1 \rangle
\end{equation}
for contracting states.

First, let us verify that this expression is equivalent to its much better-known, index-full counterpart. To this end, one expands
\begin{equation}
	\langle U | = U^{\Ih} \langle_{\Ih}|\,, \qquad \langle V | = V^{\Ih} \langle_{\Ih}|\,, \qquad \text{and} \qquad |\partial\rangle = |^{\Ih}\rangle\ \partial_{\Ih}\,,
\end{equation}
in a complete basis of states $\langle_{\Ih}|$, with $\Ih=1,\ldots,2(d+n)$, and their duals $|^{\Ih}\rangle$. We then find
\begin{equation}
	\left( \genLieM_{\langle U|} V^{\Ih} \right) \langle_{\Ih}| = \left( U^{\Jh} \partial_{\Jh} V^{\Ih} + \Zop^{\Lh\Ih}_{\Kh\Jh} V^{\Jh} \partial_{\Lh} U^{\Kh} \right) \langle_{\Ih}| \qquad \text{with} \qquad \Zop^{\Lh\Ih}_{\Kh\Jh} = \langle_{\Jh} |\langle_{\Kh}| \Zop |^{\Lh} \rangle |^{\Ih} \rangle
\end{equation}
after inserting the identity $|^{\Ih}\rangle \langle_{\Ih}| = 1$. Now, we see that
\begin{equation}\label{eqn:Zopcomponents}
	\Zop^{\Lh\Ih}_{\Kh\Jh} = \eta^{\Lh\Ih} \eta_{\Kh\Jh} - \delta_{\Jh}^{\Lh} \delta_{\Kh}^{\Ih}
\end{equation}
has to hold to obtain the expected result
\begin{equation}
	\genLieM_{\langle U|} V^{\Ih} = U^{\Jh} \partial_{\Jh} V^{\Ih} + \left( \partial^{\Ih} U_{\Jh} - \partial_{\Jh} U^{\Ih} \right) V^{\Jh} \,.
\end{equation}

%\subsection{Split-Casimir and $\Zop$ operator}\label{sec:Casimir}
We have seen that to evaluate the index-free generalized Lie derivative \eqref{eqn:genLie}, we need the operator $\Zop$ which eventually results in \eqref{eqn:Zopcomponents}. Now, we show that it can be conveniently expressed in terms of the split Casimir operator of the algebra we discussed in section~\ref{sec:algebra}.

As the name suggests, the split Casimir operator arises from the quadratic Casimir operator
\begin{equation}
	\begin{gathered}
		C = K_\alpha^\beta \cdot K_\beta^\alpha - 2 K_{AB} \cdot K^{AB} - 2 \Rt^\alpha_A \cdot R_\alpha^A - \Rt^{\alpha\beta} \cdot R_{\alpha\beta} \\
		- \kappa_{\alpha\beta} \left( \Rt^{\beta\gamma} \cdot K_\gamma{}^\alpha - \Rt^{\alpha A} \cdot \Rt^\beta_A - \tfrac{1}{4}\kappa_{\gamma\delta} \Rt^{\alpha\gamma} \cdot \Rt^{\beta\delta} \right)\,,
	\end{gathered}
\end{equation}
where we use the symmetric product
\begin{equation}
	A \cdot B = \tfrac12 ( A B + B A )\,.
\end{equation}
By construction, it commutes with all generators. Moreover, the states $|^{\Ah}\rangle$ are its eigenstates with the eigenvalue
\begin{equation}\label{eqn:eigCasimir}
	C |^{\Ah}\rangle = [ 2 ( d + n ) - 1 ] |^{\Ah}\rangle,
\end{equation}
which is exactly what one also gets from\footnote{Here the weights $\lambda(R_1)$ and $\rho$ are given in the Dynkin basis. In this basis, the pairing $(\lambda_1,\lambda_2) = \lambda_1 A^{-1} \lambda_2^T$ is given in terms of the inverse of $D_{d+n}$'s Cartan matrix $A$.}
\begin{equation}
	2 ( d +  n ) - 1 = \Bigl( \lambda(R_1) , \lambda(R_1) + 2 \rho \Bigr)
\end{equation}
with
\begin{equation}
	\lambda(R_1) = \weight{1 0 \cdots 0} \qquad \text{and} \qquad \rho = \weight{1 \cdots 1}
\end{equation}
for the $D_{d+n}$ series simple Lie algebras generating the duality group $\Odd[d+n]$ on the mega-space. Hence, we found the quadratic Casimir operator and normalized it correctly.

The split Casimir is obtained by substituting $\cdot$ with $\odot$, defined by
\begin{equation}
	A \odot B = \tfrac12 \left( A \otimes B + B \otimes A \right)\,,
\end{equation}
and resulting in
\begin{equation}
	\begin{gathered}
		\Cop = K_\alpha^\beta \odot K_\beta^\alpha - 2 K_{AB} \odot K^{AB} - 2 \Rt^\alpha_A \odot R_\alpha^A - \Rt^{\alpha\beta} \odot R_{\alpha\beta} \\
		- \kappa_{\alpha\beta} \left( \Rt^{\beta\gamma} \odot K_\gamma{}^\alpha - \Rt^{\alpha A} \odot \Rt^\beta_A - \tfrac14 \kappa_{\gamma\delta} \Rt^{\alpha\gamma} \odot \Rt^{\beta\delta} \right)\,.
	\end{gathered}
\end{equation}
We use it to eventually define
\begin{equation}
	\Zop = - \frac{\alpha}{2 I(R_1)} \Cop + \beta \, 1 \otimes 1,
\end{equation}
where the constants $\alpha$ and $\beta$ are well-known for $\Odd[d]$ and the exceptional duality groups \cite{Berman:2012vc}. Note that $I(R_1)$ is the Dynkin index of the $R_1=\text{fundamental}$ representation and is given by
\begin{equation}
	I(R) = \frac{\dim R}{2 \dim G} \left( \lambda(R), \lambda(R) + 2 \rho \right)\,.
\end{equation}
Remarkably, for all duality groups $\alpha = 2 I(R_1)$ and one therefore finds the simple relation
\begin{equation}
	\Zop = - \Cop + \beta \, 1 \otimes 1\,.
\end{equation}
Finally, $\beta=0$ for $\Odd[d]$, and the relation simplifies further to
\begin{equation}
	\Zop = - \Cop\,.
\end{equation}

By using the states defined in section~\ref{sec:vermamodule}, one can verify that
\begin{equation}\label{eqn:Zcomponentsflat}
	\Zop^{\Ah\Bh}_{\Ch\Dh} = \langle_{\Dh} |\langle_{\Ch}| \Zop |^{\Ah}\rangle |^{\Bh}\rangle = \eta^{\Ah\Bh} \eta_{\Ch\Dh} - \delta^{\Ah}_{\Dh} \delta^{\Bh}_{\Ch}
\end{equation}
holds. This is exactly the relation we derived in \eqref{eqn:Zopcomponents}, but in flat instead of curved indices. As we will discuss in section~\ref{sec:genmega-frame}, these two index types are related by the generalized frame
\begin{equation}
	\hE_{\Ah}{}^{\Ih} = \langle_{\Ah} | \hE |^{\Ih} \rangle\,,
	\qquad \text{and its inverse} \qquad
	(\hE^{-1})_{\Ih}{}^{\Ah} = \langle_{\Ih} | \hE |^{\Ah}\rangle\,.
\end{equation}
Because it is an element of the duality group $\Odd[d+n]$, one can simply swap flat for curved indices in \eqref{eqn:Zcomponentsflat} to obtain \eqref{eqn:Zopcomponents}. A quick crosscheck of this result can be done by recovering the quadratic Casimir operator $\Cop$ from $\Zop$ as
\begin{equation}
	\langle_{\Ah}| \Cop |^{\Bh} \rangle = - \langle_{\Ch}| \langle_{\Ah}| \Zop |^{\Ch} \rangle |^{\Bh} \rangle = - \Zop^{\Ch\Bh}_{\Ah\Ch} = \left[ 2(d+n) - 1 \right] \delta_{\Ah}^{\Bh}\,,
\end{equation}
which matches the expectation set by \eqref{eqn:eigCasimir}.

Finally, we want to rewrite $\Zop$ in the heterotic basis introduced in section~\ref{sec:hetBasis}. Doing so give rise to
\begin{equation}\label{eqn:ZOphetbasis}
	\begin{aligned}
		\Zop = & - \Khet_{\alphan}^{\betan} \odot \Khet_{\betan}^{\alphan} + 2 \Khet_{\Ac\Bc} \odot \Khet^{\Ac\Bc} + 2 \Rthet^{\alphan}_{\Ac} \odot  \Rhet_{\alphan}^{\Ac} + \Rthet^{\alphan\betan} \odot \Rhet_{\alphan\betan} \\
		       & - 2 \Rthet^{\alphai\betai} \odot \Rthet_{\alphai\betai} - 2 \Khet_{\alphai}^{\betan} \odot \Khet_{\betan}^{\alphai} - \Rthet^{\betai}_{\Ac} \odot \Rthet_{\betai}^{\Ac}\,.
	\end{aligned}
\end{equation}
In the first line, we collected all contributions from $\gM'$ generators, while the second line only contains generators from the set $\fp$. This particular decomposition will prove very useful in the next section.

\section{Generalized mega-frame}\label{sec:genmega-frame}
After understanding how generalized diffeomorphsims act on the mega-space, we proceed with the generalized frame that underlies the \PS{} construction. Following previous work, we choose it to have the form \cite{Butter:2022iza,Hassler:2023axp}
\begin{equation}\label{eqn:decompMegaFrame}
	\hE = \Mt \cE \Vt \qquad\text{with}\qquad \cE \in \GPS\,.
\end{equation}
Each of the components on the right-hand side is an element of the duality group $\Odd[d+n]$, while $\cE$ is further restricted to be in the subgroup $\GPS$ generated by \eqref{eqn:KABandRs}. Moreover, a tilde over a quantity emphasizes that it depends on coordinates that parameterize auxiliary directions in the mega-space. At the end of the day, we want to remove these directions to obtain the physical space. Therefore, it is in our interest to keep track of them. The objective of this section is to fix $\Mt$ and $\Vt$ completely by imposing that the mega-space torsion, which is computed with the generalized Lie derivative \eqref{eqn:genLie} by
\begin{equation}\label{eqn:genTorsion}
	\genLieM_{\langle_{\Ah}|\hE} \hE \hE^{-1} = \Th_{\Ah}\,,
\end{equation}
has certain constant components.

\subsection{Fixing \texorpdfstring{$\Mt$}{M̃}}
As a first step, we want to go from curved to flat indices. This is done by $\Vt$ with the components
\begin{equation}
	\Vt_{\Ah}{}^{\Ih} = \langle_{\Ah} | \Vt |^{\Ih} \rangle
\end{equation}
that are further restricted by
\begin{equation}\label{eqn:Vtrestrictions}
	\Vt_{\Ah}{}^{\Ih} = \begin{pmatrix}
		\Vt_\alpha{}^\mu & 0         & \Vt_{\alpha\mu}  \\
		0                & \Vt_A{}^I & 0                \\
		0                & 0         & \Vt_\mu{}^\alpha
	\end{pmatrix},
\end{equation}
with $\Vt_\alpha{}^\mu \Vt_\mu{}^\beta = \delta_\alpha^\beta$ and $\Vt_{(\alpha}{}^\mu \Vt_{\beta)\mu} = -\kappa_{\alpha\beta}$ arising from the requirement $\Vt\in\Odd[d+n]$, namely
\begin{equation}
	\eta^{\Ah\Bh} \Vt_{\Ah}{}^{\Ih} \Vt_{\Bh}{}^{\Jh} = \eta^{\Ih\Jh} = \begin{pmatrix}
		0              & 0         & \delta^\mu_\nu \\
		0              & \eta^{IJ} & 0              \\
		\delta_\mu^\nu & 0         & 0
	\end{pmatrix}\,.
\end{equation}
Applying it to the partial derivative in the definition \eqref{eqn:genLie} gives rise to a new derivative
\begin{equation}
	\Dt = \Vt \partial \,.
\end{equation}

On the mega-space, we use the solution
\begin{equation}
	\partial^\mu \,\cdot\, = 0 \,, \qquad \text{and} \qquad
	\partial_I \, \cdot\, \partial^I \,\cdot \, = 0
\end{equation}
of the section condition which is required for generalized diffeomorphisms to close. Taking into account the chosen restrictions on $\Vt$ in \eqref{eqn:Vtrestrictions}, we equivalently have
\begin{equation}
	\Dt^\alpha \, \cdot \, = 0 \qquad \text{and} \qquad \Dt_A \,\cdot\, \Dt^A \,\cdot\, = 0\,.
\end{equation}
For the tilded quantities $\Mt$ and $\Vt$ these two equations are solved by
\begin{equation}\label{eqn:solSCMtandVt}
	\Dt^\alpha \Mt = \Dt^\alpha \Vt = 0\,, \qquad \text{and} \qquad \Dt_A \Mt = \Dt_A \Vt = 0\,.
\end{equation}
Moreover, we take
\begin{equation}\label{eqn:solSCcE}
	\Dt^\alpha \cE = 0\,, \qquad
	\Dt_\alpha \cE = 0\,, \qquad
	\text{and} \qquad \Dt_A \cE \, \Dt^A \cE = 0\,,
\end{equation}
hinting that $\cE$ contains all information about the physical space, while $\Mt$ and $\Vt$ are auxiliary quantities required to tune certain components of the generalized torsion $\Th_{\Ah}$. An inconvenient property of the latter is that it still depends on the auxiliary coordinates of the mega-space. However, we can introduce the twisted torsion $T_{\Ah}$ by
\begin{equation}\label{eqn:twistedTorsion}
	\Th_{\Ah} = \Mt_{\Ah}{}^{\Bh} ( \Mt T_{\Bh} \Mt^{-1} )\,,
\end{equation}
for which this dependence will drop out -- assuming of course an appropriate choice of $\Mt$ and $\Vt$. Combining this equation with \eqref{eqn:genTorsion} and \eqref{eqn:genLie} gives rise to the compact expression
\begin{equation}\label{eqn:TAh}
	\boxed{%
	T_{\Ah} = \langle_{\Ah} | \cE |^{\Bh} \rangle \Theta_{\Bh} + \langle_{\Ah} | \Theta_{\Bh} \Zop \cE |^{\Bh}\rangle}
\end{equation}
for the twisted torsion with the Maurer-Cartan form
\begin{equation}
	\Theta_{\Ah} = \Mt^{-1} \Dt_{\Ah} \Mt + \Dt_{\Ah} \cE \cE^{-1} + \cE \Dt_{\Ah} \Vt \Vt^{-1} \cE^{-1}\,.
\end{equation}
Taking into account the restrictions \eqref{eqn:solSCMtandVt} and \eqref{eqn:solSCcE}, one finds that the only non-vanishing components of $\Theta_{\Ah}$ are
\begin{equation}\label{eqn:ThetaAhat}
	\begin{aligned}
		\Theta_\alpha & = \Mt^{-1} \Dt_\alpha \Mt + \cE \Dt_\alpha \Vt \Vt^{-1} \cE^{-1} \,, \qquad \text{and} \\
		\Theta_A      & = \Dt_A \cE \cE^{-1}\,.
	\end{aligned}
\end{equation}
Eventually, we want to fix them such that the lowest components of the twisted generalized torsion become the generators of the structure group $\GS$ we discussed in section~\ref{sec:GS}. This property lies at the heart of the \PS{} construction and is realized by imposing
\begin{equation}\label{eqn:genTorsionConstraint}
	\boxed{%
		T_\alpha = \hatt_\alpha\,.
	}
\end{equation}
For this relation to hold, we first seed $\hatt_\alpha$ by choosing
\begin{equation}
	\Mt^{-1} \Dt_\alpha \Mt = \hatt_\alpha
\end{equation}
which is equivalent to identifying $\Vt_\mu{}^\alpha \hatt_\alpha$ with the left-invariant Maurer-Cartan form
\begin{equation}\label{eqn:defVtmualpha}
	\Mt^{-1} \partial_\mu \Mt = \Vt_\mu{}^\alpha \hatt_\alpha\,.
\end{equation}
Consequentially, $\Mt$ has to be an element of the generalized structure group $\GS$.

\subsection{Fixing \texorpdfstring{$\Vt$}{Ṽ}}
What is left to check is whether the remaining component $\Vt_{\alpha\mu}$ of $\Vt$ can be fixed such that \eqref{eqn:genTorsionConstraint} holds. The central point of this computation is to make sure that the twisted torsion components $T_{\alpha}$ are independent of the choice of $\cE\in\GPS$. To show that $\cE$ decouples, we use its property
\begin{equation}
	\langle_\alpha | \cE = \langle_\alpha |\,
\end{equation}
and rewrite \eqref{eqn:TAh} as
\begin{equation}
	\cE^{-1} T_{\Ah} \cE = \langle_{\Ah} | \cE |^{\Bh}\rangle \Thetab_{\Bh} + \langle_{\Ah}| \cE \Thetab_{\Bh} \Zop |^{\Bh}\rangle,
\end{equation}
with
\begin{equation}\label{eqn:ThetaBar}
	\Thetab_{\Ah} = \cE^{-1} \Theta_{\Ah} \cE
\end{equation}
to fully exploit it. Now, we can simplify the expression of $T_\alpha$ to
\begin{align}
	T_\alpha & = \Theta_\alpha + \cE \langle_\alpha | \Thetab_\beta \Zop |^\beta\rangle \cE^{-1}                                                           \\
	         & = \hatt_{\alpha} + \cE \left( \genLieM_{\langle\Vt_\alpha|} \Vt \Vt^{-1} + \langle_\alpha| \hatt_\beta \Zop |^\beta\rangle \right) \cE^{-1}
\end{align}
and therefore fix
\begin{equation}\label{eqn:constrTalphaEqtalpha}
	\genLieM_{\langle\Vt_\alpha|} \Vt \Vt^{-1} = - \langle_\alpha| \hatt_\beta \Zop |^\beta\rangle \,.
\end{equation}
It is straightforward to evaluate the right-hand side; All one needs is given in section~\ref{sec:algebra}. Similarly for the left-hand side, but this time one also has to take into account the restrictions on $\Vt$ given in \eqref{eqn:Vtrestrictions}. Without loss of generality, they give rise to
\begin{equation}\label{eqn:DalphaVtVtinv}
	\Dt_\alpha \Vt \Vt^{-1} = w_{\alpha\beta}{}^\gamma K_\gamma^\beta - \tfrac12 w_{\alpha\beta\gamma} \Rt^{\gamma\beta},
\end{equation}
with
\begin{align}
	w_{\alpha\beta}{}^\gamma & = - \Vt_\alpha{}^\mu \Vt_{\beta}{}^\nu \partial_\mu \Vt_\nu{}^\gamma\,, \qquad \text{and}                                                              \\
	w_{\alpha\beta\gamma}    & = \phantom{-} \Vt_\alpha{}^\mu \partial_\mu \Vt_{[\beta|\nu} \Vt_{|\gamma]}{}^\nu - w_{\alpha[\beta|}{}^\delta \Vt_{\delta\mu} \Vt_{|\gamma]}{}^\mu\,,
\end{align}
under the assumption that $V_A{}^I$ is constant and therefore does not contribute. For \eqref{eqn:constrTalphaEqtalpha} to hold, we only need the components to satisfy
\begin{align}\label{eqn:wfromf}
	w_{[\alpha\beta]}{}^\gamma & = - \tfrac12 f_{\alpha\beta}{}^\gamma \qquad\text{and} \\\label{eqn:BtildeConstr}
	w_{[\alpha\beta\gamma]}    & = \phantom{-}\tfrac16 f_{\alpha\beta\gamma}\,,
\end{align}
with
\begin{equation}
	f_{\alpha\beta\gamma} = f_{\alpha\beta}{}^\delta \kappa_{\delta\gamma}\,.
\end{equation}
This tensor is totally antisymmetric because we required \eqref{eqn:kappainvariant} before. As we have already fixed $\Vt_\mu{}^\alpha$ (and with it its inverse $\Vt_\alpha{}^\mu$) by \eqref{eqn:defVtmualpha}, the first constraint does not provide any new information. Of course it still holds. Hence, the last component of $\Vt$ that has to be determined is $\Vt_{\alpha\mu}$. With out loss of generality, it can be rewritten as
\begin{equation}
	\Vt_{\alpha\mu} = -\tfrac12 \kappa_{\alpha\beta} \Vt^\beta{}_\mu + \Vt_\alpha{}^\nu \Bt_{\nu\mu}
\end{equation}
to obtain
\begin{equation}
	w_{[\alpha\beta\gamma]} = \Vt_\alpha{}^\mu \Vt_\beta{}^\nu \Vt_\gamma{}^\rho \partial_{[\mu} \Bt_{\nu\rho]}\,.
\end{equation}
Because the two-form $\Bt_{\mu\nu}$ is not invariant under generalized diffeomorphisms on the mega-space, we instead use the corresponding $H$-flux
\begin{equation}
	\Ht_{\mu\nu\rho} = 3 \partial_{[\mu} \Bt_{\nu\rho]}\,.
\end{equation}
It is constrained by \eqref{eqn:BtildeConstr} to
\begin{equation}
	\Ht_{\mu\nu\rho} = \tfrac12 f_{\alpha\beta\gamma} \Vt_\mu{}^\alpha \Vt_\nu{}^\beta \Vt_\rho{}^\gamma
\end{equation}
and leaves us with the final question, if there exists a $B$-field which gives rise to this specific $H$-flux. To answer this question in the affirmative, it is sufficient to check for
\begin{equation}
	4 \partial_{[\mu} \Ht_{\nu\rho\lambda]} = 0\,.
\end{equation}
This equation indeed holds, due to the Jacobi identity of the structure constants $f_{\alpha\beta}{}^\gamma$. Finally, we have to fix the constant $V_A{}^I$. All constant $\Odd[d]$ elements are admissible, but a particular simple choice is the identity. An immediate consequence is that
\begin{equation}
	|^A\rangle \Dt_A = |^I\rangle \partial_I\,.
\end{equation}

This completes the first major objective in exploring the twisted \PS{} construction: We have constructed the generalized frame $\hE$ on the mega-space, such that:
\begin{enumerate}
	\item Both $\Mt$ and $\Vt$ are completely fixed by requiring that $\hatt_\alpha$ arises as the component $T_\alpha = \hatt_\alpha$ of the twisted generalized torsion.
	\item The inner frame $\cE$ contains all information about the physical space. It does not depend on the auxiliary coordinates and is an element of $\GPS$.
	\item The twisted generalized torsion $T_{\Ah}$ does not depend on the auxiliary coordinates. This is the reason why we do not use the mega-space torsion $\Th_{\Ah}$ but instead its twisted version $T_{\Ah}$ which we defined in \eqref{eqn:twistedTorsion}.
\end{enumerate}

\subsection{Twisted generalized torsion}\label{sec:tgT}
By design, the twisted generalized torsion satisfies $T_\alpha = \hatt_\alpha$. Because it is not a completely free tensor, this choice also affects the other components, $T_A$, $T^\alpha$ and therefore suggests the splitting into an intrinsic and extrinsic part according to
\begin{equation}
	T_{\Ah} = \Ti_{\Ah} + \Te_{\Ah}\,.
\end{equation}
As the name implies, the intrinsic part is automatically built into the mega-space frame. It arises when we set the frame $\cE$ to the identity. Taking into account the equations from the last section, we therefore obtain
\begin{equation}
	\Ti_{\Ah} = \Thetai_{\Ah} + \langle_{\Ah}| \Thetai_\beta \Zop |^\beta \rangle
\end{equation}
with the only non-vanishing component
\begin{equation}
	\Thetai_{\alpha} = \hatt_\alpha - \tfrac12 f_{\alpha\beta}{}^\gamma K_\gamma^\beta - \tfrac1{12} f_{\alpha\beta\gamma} \Rt^{\beta\gamma}\,.
\end{equation}
of $\Thetai_{\Ah}$. Besides the already known $\Ti_{\alpha} = \hatt_\alpha$, the remaining components $\Ti_A$ and $\Ti^\alpha$ fix all contributions to the respective components of $T_{\Ah}$ for the generators $K^\alpha_\beta$, $\Rt^{\alpha}_B$ and $\Rt^{\alpha\beta}$. Therefore the remaining, free components have to go to the extrinsic torsion with $\Te_\alpha = 0$ to not conflict with $T_\alpha = \hatt_\alpha$. We parameterize it in the heterotic basis by
\begin{equation}\label{eqn:hetIntrinsic}
	\begin{aligned}
		\Te_{\Ac}     & = f_{\Ac\Bc\Cc} \Khet^{\Bc\Cc} + f_{\Ac\Bc}{}^{\gamman} \Rhet_{\gamman}^{\Bc} + \tfrac12 f_{\Ac}{}^{\betan\gamman} \Rhet_{\gamman\betan} \,, \qquad \text{and} \\
		\Te^{\alphan} & = f_{\Bc\Cc}{}^{\alphan} \Khet^{\Bc\Cc} + f_{\Bc}{}^{\gamman\alphan} \Rhet^{\Bc}_{\gamman} + \tfrac12 f^{\alphan\betan\gamman} \Rhet_{\gamman\betan}\,,
	\end{aligned}
\end{equation}
where the different $f$-tensors encode torsions and curvatures of the physical space. Therefore, we will streamline the way $\Te$ is computed in the following. Doing so, we will in particular show later that several of its components can be fixed by choosing $\cE$ appropriately.

Because all the generators that appear in the extrinsic torsion belong to $\gPS$, we can isolate them by considering the twisted torsion components $\hlangle_{\Bd} | T_{\Ah} |^{\Cd} \hrangle$. The same holds for the index $\Ah$, because we have already seen that $\Te_{\alpha} = 0$. Hence, instead of $T_{\Ah}$, we rather work with
\begin{equation}\label{eqn:restrictedTorsion}
	\THet_{\Ad} = \hlangle_{\Ad} |^{\Bh} \rangle \left. T_{\Bh} \right|_{\gPS}
\end{equation}
in the heterotic basis discussed above. The projection to $\gPS$ on the right-hand side is performed such that
\begin{equation}
	\hlangle_{\Ad} | X |_{\Bd} \hrangle =  \hlangle_{\Ad} | X |_{\gPS} |_{\Bd} \hrangle \qquad \forall\, X \in \gM
\end{equation}
holds. Practically, this means that we remove all tilded generators and $K_{\alpha}{}^\beta$ except for
\begin{align}
	\left.K_{\alphan}^{\betai}\right|_{\gPS} & = R_{\alphan}{}^{\betai}\,,          &  &            &
	\left.K_{\alphai}^{\betai}\right|_{\gPS} & = \tfrac12 R_{\alphai}{}^{\betai}\,,                   \\
	\left.\Rt^{\alphai}_B\right|_{\gPS}      & = R^{\alphai}_B\,,                   &  & \text{and} &
	\left.\Rt^{\alphai\betai}\right|_{\gPS}  & = R^{\alphai\betai}\,.
\end{align}
The resulting tensor, called the restricted twisted torsion, is denoted by $\THet_{\Ad}$. Again, it decomposes into an intrinsic and extrinsic part according to
\begin{equation}
	\THet_{\Ad} = \THeti_{\Ad} + \Te_{\Ad} \,.
\end{equation}
Remarkably, the form of the extrinsic part \eqref{eqn:hetIntrinsic} is preserved. The advantage of this form is that it can be easier computed than $T_{\Ah}$ because only the generators of $\GM'$ are involved. Hence, similar to the restriction in \eqref{eqn:restrictedTorsion}, we furthermore define $X |_{\gM'}$ to replace the generators
\begin{align}
	\left.K_{\alphan}^{\betai}\right|_{\gM'} & = R_{\alphan}{}^{\betai}\,,          & \left.K_{\alphai}^{\betan}\right|_{\gM'} & = - \tfrac12 \Rt_{\alphai}{}^{\betan}\,, &
	\left.K_{\alphai}^{\betai}\right|_{\gM'} & = \tfrac12 R_{\alphai}{}^{\betai}\,,                                                                                         \\
	\left.\Rt^{\alphai}_B\right|_{\gM'}      & = R^{\alphai}_B\,,                   &                                          & \text{and}                               &
	\left.\Rt^{\alphai\betai}\right|_{\gM'}  & = R^{\alphai\betai}
\end{align}
while keeping all the others.

To get $\THet_{\Ad}$ starting from the original expression in \eqref{eqn:TAh}, we first need some auxiliary quantities. Most important is the torsion introduced by $\Vt$ which is
\begin{equation}
	T^{\Vt}_{\Ah} = \genLieM_{\langle\Vt_{\Ah}|} \Vt \Vt^{-1}
\end{equation}
with the components
\begin{align}
	T^{\Vt}_\alpha & = -f_{\alpha\beta}{}^\gamma K_\gamma^\beta - \tfrac12 f_{\alpha\beta\gamma} \Rt^{\beta\gamma}\,, & T^{\Vt\,\alpha} & = \tfrac12 f_{\beta\gamma}{}^\alpha \Rt^{\beta\gamma} &
	T^{\Vt}_A      & = 0\,.
\end{align}
In particular, we are interested in the restricted components
\begin{equation}\label{eqn:VtorsionComponents}
	\begin{aligned}
		T^{\Vt}_{\alphan}|_{\gM'} & = - f_{\alphan\betan}{}^{\gamman} K_{\gamman}^{\betan} - \tfrac12 f_{\alphan\betan\gammai} \Rt^{\betan\gammai} - f_{\alphan\betai}{}^{\gamman} R_{\gamman}{}^{\betai}\,,                                      \\
		T^{\Vt}_{\alphai}|_{\gM'} & = - f_{\alphai\betan}{}^{\gamman} K_{\gamman}^{\betan} -\tfrac12 f_{\alphai\betan\gammai} \Rt^{\betan\gammai} - f_{\alphai\betai}{}^{\gamman} R_{\gamman}{}^{\betai}\,,                                       \\
		T^{\Vt\,\alphan}|_{\gM'}  & = \tfrac12 f_{\beta\gamma}{}^{\alphan} \Rt^{\beta\gamma} - \tfrac12 f_{\betai\gammai}{}^{\alphan} \Rt^{\betai\gammai} + \tfrac12 f_{\betai\gammai}{}^{\alphan} R^{\betai\gammai}\,,  \qquad \text{and} \qquad \\
		T^{\Vt\,\alphai}|_{\gM'}  & = \tfrac12 f^{\alphai\betai\gammai} R_{\betai\gammai} \,.
	\end{aligned}
\end{equation}
In the same vein, we define
\begin{equation}
	\htau_{\alpha} = \hatt_\alpha |_{\gPS}
\end{equation}
with the only non-vanishing component
\begin{equation}
	\htau_{\alphan} = t_{\alphan} + f_{\alphan\betai}{}^{\gamman} R_{\gamman}{}^{\betai}\,, \qquad \text{and} \qquad
	\htau_{\alphai} = t_{\alphai} - f_{\alphai\betai}{}^{\gamman} R^{\betai}{}_{\gamman} - \tfrac12 f_{\alphai\betai\gammai} R^{\betai\gammai}\,.
\end{equation}
After these preparations, we eventually obtain
\begin{equation}\label{eqn:THet}
	\THet_{\Ad} = \bcF_{\Ad} + \hlangle_{\Ad}| \cE |^\beta \rangle \htau_\beta + \hlangle_{\Ad}| \htau_\beta \ZHet \cE |^\beta \rangle\,.
\end{equation}
Several new quantities are appearing here; We explain them going from the left to the right: First, we have the generalized fluxes obtained from the twisted generalized Lie derivative
\begin{equation}\label{eqn:hetgenLie}
	\genLieHet_{\scalebox{0.75}{\hlangle} U |} \hlangle V| = \hlangle V | \hlangle U | \partial V \hrangle + \hlangle V| \hlangle U| \ZHet |\partial U\hrangle + \hlangle U|^{\Ah}\rangle T^{\Vt}_{\Ah}|_{\gM'}
\end{equation}
with the heterotic $\Zop$-operator\,,
\begin{equation}\label{eqn:ZHet}
	\ZHet = - \Khet_{\alphan}^{\betan} \odot \Khet_{\betan}^{\alphan} + 2 \Khet_{\Ac\Bc} \odot \Khet^{\Ac\Bc} + 2 \Rthet^{\alphan}_{\Ac} \odot  \Rhet_{\alphan}^{\Ac} + \Rthet^{\alphan\betan} \odot \Rhet_{\alphan\betan} \,,
\end{equation}
given by term on the first line of \eqref{eqn:ZOphetbasis}. With it, we compute the generalized fluxes
\begin{equation}\label{eqn:cFcA}
	\bcF_{\Ad} = \genLieHet_{\scalebox{0.75}{\hlangle}_{\Ad}|\cE} \cE \cE^{-1} \,.
\end{equation}
Note that $\ZHet$ arises from $\Zop$ because all generators of $\fp$ are removed through the projection $|_{\gPS}$.

The last, in \eqref{eqn:hetgenLie} describes a generalized torsion. To make contact with the heterotic string, this torsion has to have a particular form, namely
\begin{equation}\label{eqn:hetTorsion}
	\hlangle_{\Ah} | \cE |^{\Bh} \rangle T^{\Vt}_{\Bh} |_{\gPS} = \hlangle_{\Ah} | \cE |_{\betai} \hrangle \tfrac12 f^{\betai\gammai\deltai} R_{\gammai\deltai} \,.
\end{equation}
This only happens, if we impose the restriction
\begin{equation}\label{eqn:productGS}
	f_{\alphai\betai}{}^{\gamman} = 0 \,, \qquad \text{and} \qquad
	f_{\alphai\betan}{}^{\gamman} = 0 \,,
\end{equation}
Rendering the structure group $\GS=\GH\times\GS'$ into a direct product of the heterotic gauge group $\GH$, generated by $t_{\alphai}$, and the residual structure group $\GS'$ whose Lie algebra is spanned by $t_{\alphan}$. In this case, $\Vt$'s torsion components in \eqref{eqn:VtorsionComponents} simplify to
\begin{equation}\label{eqn:TVtProduct}
	\begin{aligned}
		T^{\Vt}_{\alphan}|_{\gM'} & = - f_{\alphan\betan}{}^{\gamman} K_{\gamman}^{\betan}\,,        &  &                          &
		T^{\Vt\,\alphan}|_{\gM'}  & = \tfrac12 f_{\betan\gamman}{}^{\alphan} \Rt^{\betan\gamman} \,,                                                                                                                     \\
		T^{\Vt}_{\alphai}|_{\gM'} & = 0\,,                                                           &  & \qquad \text{and} \qquad & T^{\Vt\,\alphai}|_{\gM'} & = \tfrac12 f^{\alphai\betai\gammai} R_{\betai\gammai}\,.
	\end{aligned}
\end{equation}
All components on the second line give rise to \eqref{eqn:hetTorsion}. Combined with the second term of $\ZHet$ in \eqref{eqn:ZHet}, it forms the generalized Lie derivative of heterotic double field theory with gauge group $\GH$. The terms on the first line of \eqref{eqn:TVtProduct} are of the same form as one gets for the duality group O($d$,$d$) in the standard \PS{} construction. In fact one can reuse all the expressions computed there after a slight adjustment of the indices. Another way to obtain this result is to remember that because of the product form of $\GS$, one can do the reduction from the mega-space to the physical space in two steps: First we eliminate the auxiliary coordinates of the heterotic gauge group $\GH$ with the \PS{} construction for a non-degenerate $\kappa_{\alphai\betai}$ presented here. The resulting heterotic frame is further reduced by the standard \PS{} construction which indices $\alpha$ substituted by $\alphan$ and $A$ by $\Ac$.

We see that the mega-space frame $\cE$ contributes in two different ways to the restricted twisted torsion in \eqref{eqn:THet}: All terms coming from the generalized Lie derivative, except for the torsion, contain one derivative. Whereas the frame enters through the last two terms algebraically. The form of the differential part is very similar to the O($d$,$d$) case. Additionally, the algebraic part gives an interesting glimpse at how to fix additional components of the mega-space frame which go beyond the frame on the physical space. This is not of immediate concern when discussing generalized dualities because they use the extended space as starting point. However, one might also want to turn the argument around and try to reconstruct the mega-space for general backgrounds. In this case, one needs to impose constraints on the restrict generalized torsion in such a way that they ideally only leave the physical frame $\cE$ as a free parameters.

Assuming the factorization of the generalized structure group implied by \eqref{eqn:productGS}, we should think more about how the generators $\hatt_\alpha$ should be chosen. The simplest possibly is to just set
\begin{equation}\label{eqn:tauProductCanonical}
	\htau_{\alphai} = 0
\end{equation}
which in combination with \eqref{eqn:productGS} implies
\begin{equation}\label{eqn:hattalphaiHet}
	\hatt_{\alphai} = K_{\alphai} + \tfrac12 f_{\alphai\betai\gammai} R^{\betai\gammai} = f_{\alphai\betai\gammai} \Rthet^{\betai\gammai}\,,
\end{equation}
and is therefore compatible with \eqref{eqn:commtalphabeta}. This can be easily seen by remembering that the generators $\Rthet^{\beta\gamma}$ introduced in \eqref{eqn:genIPRthetii} generate O($p$,$q$). Therefore, the heterotic gauge group $\GH$ is embedded into this group by the structure constants $f_{\alphai\betai}{}^{\gammai}$. In this case, the components of the reduced, twisted torsion are
\begin{equation}\label{eqn:locking1}
	\THet_{\Ad\Bd\Cd} = \hlangle_{\Bd} | \THet_{\Ad} |_{\Cd} \hrangle =  \cF_{\Ad\Bd\Cd} + 3 \cE_{[\Ad|}{}^{\deltan} (\htau_{\deltan})_{|\Bd\Cd]},
\end{equation}
with $\cF_{\Ad\Bd\Cd} = \hlangle_{\Bd} | \cF_{\Ad} |_{\Cd} \hrangle$, $\cE_{\Ad}{}^{\betan} = \hlangle_{\Ac} | \cE |^{\betan} \hrangle$, and $(\tau_{\alphan})_{\Bd\Cd} = \hlangle_{\Bd} | \tau_{\alphan} |_{\Cd} \hrangle$. Remarkably, \eqref{eqn:locking1} has the same form as it would have in the \PS{}-construction for O($d$,$d$). It would be desirable to keep this from after transitioning to more complicated realizations for $\htau_{\alphai}$. This possible by shifting the additional term in \eqref{eqn:locking1} into the heterotic generalized Lie derivative on the physical space. However, it is usually preferable to work with untwisted generalized Lie derivatives because they describe the standard symmetries of the physical fields. Hence, one aims to shift the twist from the Lie derivative to the frame. We will show how this works for the example in section~\ref{sec:cigarToTrumpet}.

Fixing at least parts of components $\cE_{\Ad}{}^{\betan}$ systematically requires more information about the structure group $\GS'$. However, one important lesson which can be already learned here is that the inner frame $\cE$ should be split according to
\begin{equation}\label{eqn:cEtocA}
	\boxed{\GPS \ni \cE = \cA E \qquad \text{with} \qquad E \in \mathrm{O}(d+p,d+q)}
\end{equation}
into two parts. The first part $\cA$, is valued in the coset $\GPS/\mathrm{O}(d+p,d+q)$ and should be constraint through torsion constraints such that ideally only the physical frame $E$ -- which is generated by $\Khet_{\Ac\Bc}$ -- survives as it contains all the physical degrees of freedom. This works because by construction $E |^{\alpha}\hrangle = |^{\alpha}\hrangle$ holds and we therefore are left with
\begin{equation}
	\cE_{\Ad}{}^{\betan} = \hlangle_{\Ad}| \cE |^{\betan} \hrangle = \hlangle_{\Ad} | \cA |^{\betan} \hrangle = \cA_{\Ad}{}^{\betan}\,.
\end{equation}
Hence, it is possible to replace $\cE$ with $\cA$ in \eqref{eqn:locking1}. We will introduce an explicit parameterization of $\cA$ later and then compute the relevant components of the restricted twisted torsion.

\subsection{Dilatonic torsion}\label{sec:dilatontorsion}
Last but not least, we have to lift the dilaton to the mega-space. It is captured by the dilatonic torsion
\begin{equation}\label{eqn:defThatA}
	| \Th \rangle = 2 \hE | \partial \Phih \rangle  - \partial_{\Ih} \hE |^{\Ih}\rangle\,,
\end{equation}
which corresponds to the one-index generalized flux of double field theory's flux formulation \cite{Geissbuhler:2013uka}. Like before, we twist it to remove all dependence on auxiliary coordinates and obtain
\begin{equation}\label{eqn:defKetX}
	| T \rangle = \Mt^{-1} | \Th \rangle = 2 \cE | \Dt \Phih \rangle - \Mt^{-1} \Dt_{\Ah} ( \Mt \cE \Vt) \Vt^{-1} |^{\Ah} \rangle,
\end{equation}
with the derivative
\begin{equation}
	| \Dt \Phih \rangle = \Dt_A \Phih |^A\rangle + \Dt_\alpha \Phih |^\alpha\rangle
\end{equation}
of the generalized dilaton $\Phih$. Although Pol\'a\v{c}ek and Siegel did not discuss the dilaton in \cite{Polacek:2013nla}, we follow the same idea \cite{Butter:2021dtu,Butter:2022iza} and define the \PS{} form of the mega-space dilaton
\begin{equation}\label{eqn:dilatonPSform}
	\Phih = \frac12 \left( \mt + \log\det ( \Vt_\alpha{}^\mu) \right) + \Phi
\end{equation}
in terms of the generalized dilaton $\Phi$ on the physical space. The first term on the right hand side only depends on the auxiliary coordinates. Its $\mt$ will be chosen to fix
\begin{equation}
	\boxed{%
		\langle_\alpha | T\rangle = - \langle_\alpha| \hatt_\beta |^\beta\rangle = f_{\alpha\beta}{}^\beta\,,}
\end{equation}
the dilatonic version of \eqref{eqn:genTorsionConstraint}. It will source the intrinsic dilatonic torsion in analogy to the discussion of the last subsection. Let us see how this works out by first computing
\begin{equation}
	\langle_\alpha | T \rangle = 2 \Dt_\alpha \Phih - f_{\beta\alpha}{}^\beta - \langle_\alpha | \Dt_{\Ah} \Vt \Vt^{-1} |^{\Ah} \rangle\,.
\end{equation}
To evaluate the last term on the right-hand side, we remember \eqref{eqn:DalphaVtVtinv} and find
\begin{equation}
	\Dt_{\Ah} \Vt \Vt^{-1} |^{\Ah} \rangle = w_{\beta\alpha}{}^\beta |^\alpha\rangle\,,
\end{equation}
which eventually gives rise to
\begin{equation}
	\langle_\alpha | T \rangle = 2 \Dt_\alpha \Phih - f_{\beta\alpha}{}^{\beta} - w_{\beta\alpha}{}^{\beta} = 2 \Dt_\alpha \Phih - w_{\alpha\beta}{}^\beta\,,
\end{equation}
where \eqref{eqn:wfromf} was used in the last step. For the mega-space dilaton in \eqref{eqn:dilatonPSform}, the right-hand side becomes
\begin{equation}
	\langle_\alpha | T \rangle = \Dt_{\alpha} \mt = f_{\alpha\beta}{}^{\beta}\,,
\end{equation}
due to
\begin{equation}
	\Vt_\alpha{}^\mu \partial_\mu \log\det( \Vt_\beta{}^\nu ) = \Vt_\alpha{}^\mu \partial_\mu \Vt_\beta{}^\nu \Vt_\nu{}^\beta = w_{\alpha\beta}{}^\beta
\end{equation}
and $\Dt_\alpha \Phi = 0$. As result, we obtain the differential equation
\begin{equation}
	\dd \mt = \Vt_\mu{}^\alpha f_{\alpha\beta}{}^\beta \dd x^\mu\,.
\end{equation}
It can be integrated at least locally (and this is all we need) due to the Jacobi identity for the structure constants $f_{\alpha\beta}{}^\gamma$. With this result, we are able to bring \eqref{eqn:defKetX} it into its final form,
\begin{equation}\label{eqn:twistedDilatonTorsionFinal}
	| T \rangle = F_{\Ah} |^{\Ah} \rangle - \hatt_\alpha \cE |^\alpha\rangle\,,
\end{equation}
which further takes into account the parameterization of $\cE$ in \eqref{eqn:cEtocA} and makes use of the one-index generalized flux
\begin{equation}
	F_{\Ah} = 2 \cE_{\Ah}{}^I \partial_I \Phi - \partial_I \cE_{\Ah}{}^I
\end{equation}
on the physical space.

Already without going to the heterotic basis, \eqref{eqn:twistedDilatonTorsionFinal} has the same form as \eqref{eqn:THet} from the previous section. The first term on its right-hand side vanishes if we take $\cE$ to be constant, for example as the identity. Hence, it only contributes to the extrinsic dilatonic torsion. As before, we obtain the physical content of the dilatonic torsion by projecting it to $\hlangle_{\Ad}|$, giving rise to
\begin{equation}\label{eqn:locking2}
	\THet_{\Ad} = \hlangle_{\Ad} | T \rangle = F_{\Ad} - \hlangle_{\Ad} | \htau_\alpha \cE |^\alpha \rangle\,,
\end{equation}
with the one-index generalized flux
\begin{equation}
	F_{\Ad} = 2 \cE_{\Ad}{}^I \partial_I \Phi - \partial_I \cE_{\Ad}{}^I
\end{equation}
on the mega-space. Assuming moreover that the generalized structure group decomposes into the direct product $\GS = \GH \times \GS'$ and  \eqref{eqn:tauProductCanonical} holds, \eqref{eqn:locking2} simplifies further to
\begin{equation}
	\THet_{\Ad} = \hlangle_{\Ad} | T \rangle = F_{\Ad} - \hlangle_{\Ad} | \htau_{\alphan} \cE |^{\alphan} \hrangle\,.
\end{equation}

\subsection{Generalized structure group transformations}
A key feature of the \PS{} construction is that the twisted torsion we discussed above provides various covariant tensors under the action of $\GS$ and generalized diffeomorphisms on the physical space. These tensors will become crucial in constructing an invariant action in section~\ref{sec:PSaction}. In this section, we discuss the corresponding transformations.

More specifically, we are interested in generalized diffeomorphisms
\begin{equation}\label{eqn:genDiffEh}
	\delta_\xi \hE = \genLieM_{\langle \xih |} \hE
\end{equation}
on the mega-space that preserve the form of $\hE$ introduced in \eqref{eqn:decompMegaFrame}. Therefore, $\Mt$ and $\Vt$ should not be affected by them. To see which generalized diffeomorphisms are admissible, a convenient quantity to analyze is the twisted shift of the frame
\begin{equation}
	\delta_\xi \mathbb{E} = \Ac^{-1} \Mt^{-1} (\delta_\xi \hE) \Vt^{-1} E^{-1} = \Ac^{-1} \delta_\xi \cA + \delta_\xi E E^{-1}
\end{equation}
because it only contains the dynamic parts on the right-hand side. Using the definition of the generalized Lie derivative \eqref{eqn:genLie} together with \eqref{eqn:genDiffEh}, one finds that it is given by the expression
\begin{equation}
	\delta_\xi \mathbb{E} = \xi^{\Ah} \cA^{-1} \Theta_{\Ah} \cA + \langle \xi | \Dt_{\Ah} (E \Vt ) (E \Vt)^{-1} \Zop |^{\Ah}\rangle + \Dt_{\Ah} \xi^{\Bh} \langle_{\Bh}| \Zop E |^{\Ah}\rangle,
\end{equation}
with
\begin{equation}
	\langle \xi | = \xi^\alpha \langle_\alpha| + \xi^A \langle_A | + \xi_\alpha \langle^\alpha |\,,
	\qquad \text{and} \qquad
	\langle \xih | = \langle \xi | E \Vt\,.
\end{equation}
This equation can be simplified to
\begin{equation}\label{eqn:deltaE1}
	\delta_\xi \mathbb{E} = \genLieM_{\langle\xih |} (E \Vt) (E \Vt)^{-1} + \xi^\alpha \cA^{-1} \hatt_\alpha \cA + \xi^{\Bc} \cA^{-1} D_{\Bc} \cA + \Dt_{\Ah} \xi^{\Bh} \langle_{\Bh}| \Zop E |^{\Ah}\rangle\,.
\end{equation}
Note that we have introduced the flat derivative on the physical space,
\begin{equation}
	D_{\Ac} = E_{\Ac}{}^I \partial_I \qquad \text{with} \qquad E_{\Ac}{}^I = \langle_{\Ac} | E |^I \rangle\,
\end{equation}
in this equation. By definition, $\delta\mathbb{E}$ can only contain generators of $\GPS$. This is already true for the two terms in the middle of \eqref{eqn:deltaE1}, while the first and the last term require additional care. To eliminate any contributions from $K_\alpha^\beta$, $\Rt^\alpha_A$ and $\Rt^{\alpha\beta}$ introduced by them, we require
\begin{equation}
	\xi_\alpha = \kappa_{\alpha\beta} \xi^\beta
	\qquad \text{and} \qquad
	\Dt_\alpha \xi^{\Bh} = 0\,,
\end{equation}
resulting in
\begin{equation}\label{eqn:braxi}
	\langle \xi | = \xi^\alpha \left( \langle_\alpha | + \kappa_{\alpha\beta} \langle^\beta| \right) + \xi^A \langle_A |
\end{equation}
or written in the heterotic basis
\begin{equation}
	\hlangle \xi | = \xi^{\cA} \hlangle_{\cA} | + \xi^{\alphan} \hlangle_{\alphan} | \,.
\end{equation}
We can now further simplify the right-hand side of \eqref{eqn:deltaE1} to get
\begin{equation}\label{eqn:gaugePreMaster}
	\begin{aligned}
		\delta_\xi \mathbb{E} = & \left( 2 D_{[\Bc} \xi_{\Ac]} +  \xi^{\Cc} F_{\Cc\Ac\Bc} \right) \Khet^{\Ac\Bc} + \xi^{\Bc} \cA^{-1} D_{\Bc} \cA                                                                     \\
		                        & - D_{\Ac} \xi^{\betan} R_{\betan}^{\Ac} - \xi^{\alpha} \left( K_{\alpha} + \tfrac12 f_{\alpha\betai\gammai} R^{\betai\gammai} \right) + \xi^{\alpha} \cA^{-1} \hatt_{\alpha} \cA\,,
	\end{aligned}
\end{equation}
with the generalized fluxes from \eqref{eqn:FAcBcCc}. To further split the gauge indices in the last two terms, we assume a product generalized structure group $\GS=\GH\times\GS'$ and \eqref{eqn:hattalphaiHet}. Taking the latter into account, we see that $\cA$ commutes with $\hatt_{\alphai}$ and therefore
\begin{equation}
	\xi^{\alphai} \cA^{-1} \hatt_{\alphai} \cA - \xi^{\alphai} \left( K_{\alphai} + \tfrac12 f_{\alphai\betai\gammai} R^{\betai\gammai} \right) =
	\xi^{\alphai} \hatt_{\alphai} - \xi^{\alphai} \left( K_{\alphai} + \tfrac12 f_{\alphai\betai\gammai} R^{\betai\gammai} \right) = 0
\end{equation}
vanishes. Hence we eventually find
\begin{equation}\label{eqn:gaugetrMaster}
	\delta_\xi \mathbb{E} = \left( 2 D_{[\Bc} \xi_{\Ac]} +  \xi^{\Cc} F_{\Cc\Ac\Bc} \right) \Khet^{\Ac\Bc} + \xi^{\Bc} \cA^{-1} D_{\Bc} \cA   - D_{\Ac} \xi^{\betan} R_{\betan}^{\Ac} - \xi^{\alphan} K_{\alphan} + \xi^{\alphan} \cA^{-1} \hatt_{\alphan} \cA\,.
\end{equation}

The first term of this transformation captures undeformed generalized diffeomorphisms on the physical space. To see why, note that
\begin{equation}
	\langle_{\Ac} | \delta_\xi E E^{-1} |_{\Bc} \rangle = \delta_\xi E_{\Ac}{}^{\Ic} E_{\Ic\Bc} = \genLie_\xi E_{\Ac}{}^{\Ic} E_{\Bc\Ic} = 2 D_{[\Bc} \xi_{\Ac]} + 3 \xi^{\Cc} F_{\Cc\Ac\Bc}
\end{equation}
holds if we only focus on the contribution from $\xi^{\Ac}$ while setting $\xi^{\alphan} = 0$. Thus, the physical frame $E_{\Ac}{}^{\Ic}$ transforms exactly as it should in heterotic double field theory with the gauge group $\GH$ specified by the structure constants $f_{\alphai\betai}{}^{\gammai}$. Moreover, the second term of \eqref{eqn:gaugetrMaster} tells us that all the other fields, no matter how they particularly enter into $\cA$, have to transform as (generalized) scalars. This makes perfect sense because neither of them has any curved indices. We conclude that the action of generalized diffeomorphisms is generated by the component $\xi^{\Ac}$. It is not modified and therefore does not require further attention. The situation is more involved for structure group transformations, generated by $\xi^{\alphan}$. To see their action in more detail, one can choose an explicit parameterization for $\cA$ as we will do in the next section.

The same logic, we have to follow for the generalized dilaton on the mega-space. In analogy with \eqref{eqn:genDiffEh}, we have now
\begin{equation}
	\delta_\xi \Phih = \genLieM_{\langle \xih |} \Phih = \xi^\mu \partial_\mu \Phih - \tfrac12 \partial_\mu \xi^\mu + \xi^I \partial_I \Phi - \tfrac12 \partial_I \xi^I\,.
\end{equation}
Remarkably, the first two terms cancel for the generalized dilaton we have chosen. This can be seen from
\begin{equation}
	\xi^\mu \partial_\mu \Phih - \tfrac12 \partial_\mu \xi^\mu =
	\xi^\alpha \Dt_\alpha \Phih - \tfrac12 \xi^\alpha \partial_\mu \Vt_\alpha{}^\mu = \xi^\alpha \left( \tfrac12 f_{\alpha\beta}{}^\beta + \tfrac12 w_{\alpha\beta}{}^\beta - \tfrac12 w_{\beta\alpha}{}^\beta \right) = 0\,.
\end{equation}
Consequentially, $\Phi$ indeed transforms as
\begin{equation}
	\delta_\xi \Phi = \xi^I \partial_I \Phi - \tfrac12 \partial_I \xi^I\,,
\end{equation}
the standard transformation of the generalized dilaton on the physical space.

There are two other fundamental objects for which we would like to derive the transformation, the twisted torsion $T_{\Ac}$ and its dilatonic counterpart $| T \rangle$. Let us begin with the former whose transformation is given by
\begin{equation}
	\delta_\xi T_{\Ah} =  \xi^B D_B T_{\Ah} + \xi^\gamma \langle_{\Ah}| \hatt_\gamma |^{\Bh} \rangle T_{\Bh} + \xi^\gamma [ \hatt_\gamma, T_{\Ah} ]\,.
\end{equation}
It is not surprising that the first term again describes generalized diffeomorphisms for a scalar on the physical space. Much more interesting are the second and third terms. Hence, we restrict the discussion to $\xi^{\Ac}=0$ and deal with
\begin{equation}\label{eqn:deltaTAh}
	\delta_\xi T_{\Ah} = \xi^{\gamman} \langle_{\Ah}| \hatt_{\gamman} |^{\Bh} \rangle T_{\Bh} + \xi^{\gamman} [ \hatt_{\gamman}, T_{\Ah} ]
	\qquad \text{for} \qquad \xi^{\Ac} = 0\,.
\end{equation}
As we have seen in section~\ref{sec:tgT}, $T_{\Ah}$ splits into two contributions, the intrinsic and the extrinsic torsion. Under the transformation \eqref{eqn:deltaTAh} the latter, which contains all information about the physical space, transforms into itself. Saying that if we start from a torsion contribution in $\gPS$, it will stay in this subspace after the transformation. For the intrinsic torsion, the situation is more subtle because it might bleed into $\gPS$ and therefore mix with the extrinsic torsion. Such a mixing results in a shift symmetry which usually is not desirable as it competes with torsion constraints. To avoid it, one imposes
\begin{equation}
	\delta_\xi \THeti_{\Ad} |_{\gPS} = \xi^{\gamman} \left( \hlangle_{\Ad} | \hatt_{\gamman} |^{\Bc} \hrangle \hatt_{\Bc} + \hlangle_{\Ad} | \hatt_{\gamman} |^{\betan} \hrangle \hatt_{\betan} + [ \hatt_{\gamman}, \hatt_{\Ad} ]\right)|_{\gPS} = 0\,.
\end{equation}
Here, we have introduced the new generators $\hatt_{\Ah}$ by
\begin{equation}
	\hatt_{\Ad} = \hlangle_{\Ad} |^{\Bh}\rangle \Ti_{\Bh}
\end{equation}
in analogy with $\hatt_{\alpha} = \Ti_{\alpha}$. This constraint is obviously solved by
\begin{equation}
	[ \hatt_{\alphan}, \hatt_{\Bd} ] |_{\gPS} = - (t_{\alphan})_{\Bd}{}^{\Cc} \hatt_{\Cc} - (t_{\alphan})_{\Bd}{}^{\gamman} \hatt_{\gamman} |_{\gPS} \,.
\end{equation}
For the dilatonic torsion, the situation is similar. The mega-space transformation takes the simple form
\begin{equation}
	\delta_\xi | T \rangle = \xi^A D_A | T \rangle + \xi^\alpha \hatt_\alpha | T \rangle\,.
\end{equation}

\section{Invariant action and parameterization of \texorpdfstring{$\cA$}{A}}\label{sec:PSaction}
After understanding how the symmetries of the physical theory arise from the mega-space, we proceed with the construction of an invariant action. Ideally it should be formulated on the mega-space because all symmetries are manifest there. In principle, any combination of invariant terms are allowed. Even when we restrict the discussion to expressions with only two derivatives, there are still various possibilities. To construct a unique two derivative action we should remember that we have introduced auxiliary fields contained in $\cA$ which in principle should not appear as physical degrees of freedom. The mega-space action should be constructed such that they either do not appear at all or can be integrated out. In particular, this implies that their equations of motion should be algebraic.

To this end, consider the action
\begin{equation}
	\begin{aligned}
		S = \int \dd^{d+n} x \, e^{-2\Phih} \Bigl( & \cH^{\Ah\Bh} \Dh_{\Ah} \Th_{\Bh} + \tfrac12 \cH^{\Ah\Bh} \Th_{\Ah}\Th_{\Bh} - \tfrac14 \cH^{\Ah\Dh} \Th_{\Ah\Bh\Ch} \Th_{\Dh}{}^{\Bh\Ch} \\
		                                           & + \tfrac1{12} \cH^{\Ah\Bh} \cH^{\Ch\Dh} \cH^{\Eh\Fh} \Th_{\Ah\Ch\Eh} \Th_{\Bh\Dh\Fh}\Bigr)\,.
	\end{aligned}
\end{equation}
Because $\cH$ commutes with the action of the generalized structure group, it is invariant under $\GS$-transformations. Taking the explicit from of $\cH^{\Ah\Bh}$ given in \eqref{eqn:genMetricStandard}, this action simplifies to
\begin{equation}\label{eqn:PSaction}
	\begin{aligned}
		S  = V \int \dd^d x \, e^{-2\Phi} \Bigl( & \cH^{\Ac\Bc} \nabla_{\Ac} \THet_{\Bc} + \tfrac12 \cH^{\Ac\Bc} \THet_{\Ac} \THet_{\Bc} - \tfrac14 \cH^{\Ac\Dc} \THet_{\Ac\Bc\Cc} \THet_{\Dc}{}^{\Bc\Cc}                             \\
		                                         & + \tfrac1{12} \cH^{\Ac\Bc} \cH^{\Cc\Dc} \cH^{\Ec\Fc} \THet_{\Ac\Cc\Ec} \THet_{\Bc\Dc\Fc} - \tfrac14 \cH^{\Ac\Dc} \THet_{\Ac\Bc}{}^{\gamman} (\htau_{\gamman})_{\Dc}{}^{\Bc} \Bigr)
	\end{aligned}
\end{equation}
after expanding the indices and integration out the auxiliary directions which just contribute with a finite volume factor $V$ for compact structure groups. Note that we use here the covariant derivative defined by
\begin{equation}
	\Dh_{\Ah} \left( \Mt_{\Bh}{}^{\Ch} V_{\Ch} \right) = \Mt_{\Ah}{}^{\Bh} \Mt_{\Ch}{}^{\Dh} \nabla_{\Ch} V_{\Dh}\,,
\end{equation}
and accordingly for higher rank tensors, as it was introduced in \cite{Butter:2022iza}.

Although not obvious at the moment, we will see that this action does not receives any contributions from $\cA$. A straightforward way to see this, is to choose an explicit parameterization like
\begin{equation}
	A = \exp\left( \Omega^{\alphan}_{\Ac} \Rhet^{\Ac}_{\alphan} + \tfrac12 \rho^{\alphan\betan} \Rhet_{\alphan\betan} \right)\,.
\end{equation}
to compute
\begin{align}
	\THet_{\Ac}                & = F_{\Ac} - \Omega^{\Bc}{}_{\Bc\Ac} \,                                                                                                                                                                                                                                                                                                                     \\
	\THet_{\Ac\Bc\Cc}          & = F_{\Ac\Bc\Cc} - 3 \Omega_{[\Ac\Bc\Cc]}\,                                                                                                                                                                                                                                                                                                                 \\
	\THet_{\Ac\Bc}{}^{\gamman} & = 2D_{[\Ac}\Omega_{\Bc]}^{\gamman} - \Omega_{\Ec}^{\gamman}F^{\Ec}{}_{\Ac\Bc} -\left(\rho^{\gamman\deltan}+\frac{1}{2}\Omega^{\gamman}_{\Ec} \Omega^{\deltan \Ec}\right) (\htau_{\deltan})_{\Ac\Bc} - 2\Omega_{[\Ac}^{\deltan} (\htau_{\deltan})_{\Bc]}{}^{\gamman} - \Omega_{\Ac}{}^{\deltan} \Omega_{\Bc}{}^{\epsilonn} f_{\deltan\epsilonn}{}^{\gamman}
\end{align}
where $\Omega_{\Ac\Bc\Cc} = -\Omega_{\Ac}^{\gamman}(\htau_{\gamman})_{\Bc\Cc}$, and from the last expression
\begin{equation}
	\begin{aligned}
		\THet_{\Ac\Bc\Cc\Dc} & = \THet_{\Ac\Bc}{}^{\epsilonn} (\htau_{\epsilonn})_{\Cc\Dc}                                                                                                                                                        \\
		                     & = -2D_{[\Ac}\Omega_{\Bc]\Cc\Dc} + [\Omega_{\Ac},\Omega_{\Bc}] + \THet_{\Ac\Bc}{}^{\Ec}\Omega_{\Ec\Cc\Dc} + 2\Omega_{[\Ac\Bc]}{}^{\Ec}\Omega_{\Ec\Cc\Dc} + \frac{1}{2} \Omega^{\Ec}{}_{\Ac\Bc} \,\Omega_{\Ec\Cc\Dc} \\
		                     & \quad  +\rho_{\Ac\Bc;\Cc\Dc} - 2\Omega_{[\Ac}^{\deltan} (\htau_{\deltan})_{\Bc]}{}^{\gamman} (\htau_{\gamman})_{\Cc\Dc}
	\end{aligned}
\end{equation}
follows after defining $\rho_{\Ac\Bc;\Cc\Dc} = \rho^{\alphan\betan}(\htau_{\alphan})_{\Ac\Bc}(\htau_{\betan})_{\Cc\Dc}$. It is not surprising that they formally agree with the results found in \cite{Butter:2022iza}, as setting $\kappa_{\alpha\beta} = 0$ reproduces the O($d$,$d$) setting discussed there. Plugging them into the action \eqref{eqn:PSaction}, we recover the action of heterotic double field theory \eqref{eqn:ShetDFT}. In particular, all terms involving components for $\cA$, like $\Omega^{\alphan}_{\Ac}$ and $\rho^{\alphan\betan}$, have dropped out.

We note that the gauge transformation \eqref{eqn:gaugetrMaster} of $\Omega_{\Ac\Bc\Cc}$ can be found as
\begin{equation}
	\begin{aligned}
		\delta_\xi \Omega_{\Ac\Bc\Cc} & = \xi^{\Dc} D_{\Dc} \Omega_{\Ac\Bc\Cc} - 3\,\xi_{[\Ac|}{}^{\Dc}\,\Omega_{\Dc|\Bc\Cc]} + D_{\Ac}\xi_{\Bc\Cc} - \xi^{\alphan} (t_{\alphan})_{\Ac}{}^{\gamman} (\htau_{\gamman})_{\Bc\Cc}\,,
	\end{aligned}
\end{equation}
where $\xi_{\Ac\Bc} = \xi^{\gamman}(\htau_{\gamman})_{\Bc\Cc}$ and the transformation of $\rho_{\Ac\Bc;\Cc\Dc}$ can also be found similarly.

\section{From the cigar to the trumpet}\label{sec:cigarToTrumpet}
To illustrate the construction we describe in this paper, let us consider an example based on the coset mega-space SU(2)$\times$SU(2)/SU(2)$_\mathrm{diag}$. Despite its minimalistic character, this setup is sufficient to demonstrate most of what was discussed in the previous sections. Moreover, it is a role model on how to obtain a large class of new generalized dualities for heterotic/type I strings. As it is common in this context, we start with fixing the algebra and from there construct the physical fields.

Generators of each factor in SU(2)$\times$SU(2) are denoted by $t_{\aL}$ and $t_{\aR}$ respectively. They are governed by the commutation relations
\begin{equation}
	[t_{\aL}, t_{\bR}] = 0\,, \qquad
	[t_{\aL}, t_{\bL}] = - \epsilon_{\aL\bL}{}^{\cL} t_{\cL}\,,
	\qquad \text{and} \qquad
	[t_{\aR}, t_{\bR}] = - \epsilon_{\aR\bR}{}^{\cR} t_{\cR}
\end{equation}
where $\epsilon$ is the totally antisymmetric tensor with $\epsilon_{12}{}^3 = 1$. Moreover, we have the invariant pairing
\begin{equation}
	\llangle t_{\aL} , t_{\bL} \rrangle = \delta_{\aL\bL}\,, \qquad
	\llangle t_{\aR} , t_{\bR} \rrangle = - \delta_{\aR\bR}\,,
	\qquad\text{and}\qquad
	\llangle t_{\aL} , t_{\bR} \rrangle = 0\,.
\end{equation}
As the generalized structure group, we choose a U(1) generated by
\begin{equation}
	t_{\onei} = \tfrac{1}{\sqrt{2}} \left( t_{\oneL} + \lambda t_{\oneR} \right)
\end{equation}
which gives rise to
\begin{equation}
	\llangle t_{\onei}, t_{\onei} \rrangle = \tfrac12 ( 1 - \lambda^2 ) = - \kappa_{\onei\onei}\,.
\end{equation}
$\lambda$ has been introduced to be able to tune the $\kappa_{\alphai\betai}$. There are two special points, namely $\lambda = \pm 1$, where the invariant metric vanishes and one recovers the original \PS{} construction. Furthermore, we choose
\begin{equation}\label{eqn:tu1i}
	t^{\onei} = \frac{\sqrt{2}}{1\pm\lambda} \left( t^{\oneL} \mp t^{\oneR} \right)
\end{equation}
to guarantee the additional pairings
\begin{equation}
	\llangle t^{\onei}, t_{\onei} \rrangle = 1\,,
	\qquad \text{and} \qquad
	\llangle t^{\onei}, t^{\onei} \rrangle = 0\,.
\end{equation}
Note that according to the sign choice in \eqref{eqn:tu1i} corresponds  $\lambda = \pm 1$ are excluded. By treating these two cases separately the parameter $\lambda$ covers the full real line. Finally, we fix
\begin{equation}
	t_a = \tfrac{1}{\sqrt{2}} \left( t_{\aL} - t_{\aR} \right)\,,
	\qquad \text{and} \qquad
	t^a = \tfrac{1}{\sqrt{2}} \left( t_{\aL} + t_{\aR} \right)
\end{equation}
to have a full set of generators and with them all algebraic information fixed.

On the mega-space, we now construct a SU(2)$\times$SU(2) generalized frame for the coset element
\begin{equation}
	m = \begin{cases}
		e^{\sqrt{2} x^2 t_2} \, e^{\sqrt{2} x^1 t_1}\, e^{\sqrt{2} y t_{\onei}} \\
		e^{\sqrt{2} x^2 t_2} \, e^{\sqrt{2} x^1 t^1}\, e^{\sqrt{2} y t_{\onei}}
	\end{cases}
\end{equation}
for each sign choice in \eqref{eqn:tu1i}, following for example the prescription outlined in \cite{Borsato:2021vfy}. The result can be brought in the standard form given in \eqref{eqn:decompMegaFrame} with
\begin{equation}
	\hlangle_{\Ac} | \cE |^{\Bc} \hrangle = \begin{cases} \frac{1}{\sqrt{2}} \begin{pmatrix}
			                   s_1 t_2 \quad                                      & c_1 \quad & 0  \quad            & 0 \quad     & 0                            \\
			                   \phantom{-}\lambda_- \lambda^{-1}_+ c_1 ct_2 \quad & 0   \quad & - 2 c_1 ct_2  \quad & 2 s_1 \quad & \sqrt{2} \lambda_- c_1 ct_2  \\
			                   - \lambda_- \lambda^{-1}_+ s_1 ct_2 \quad          & 0 \quad   & 2 s_1 ct_2          & 2 c_1 \quad & -\sqrt{2} \lambda_- s_1 ct_2 \\
			                   - c_1 t_2 \quad                                    & s_1 \quad & 0 \quad             & 0 \quad     & 0                            \\
			                   2 \lambda_+^{-1} \quad                             & 0 \quad   & 0 \quad             & 0 \quad     & \sqrt{2}
		                   \end{pmatrix}
		\\
		\frac{1}{\sqrt{2}} \begin{pmatrix}
			                   - s_1 ct_2 \quad                                  & c_1 \quad & 0  \quad          & 0 \quad     & 0                           \\
			                   c_1 ct_2 \quad                                    & s_1 \quad & 0  \quad          & 0 \quad     & 0                           \\
			                   \phantom{-}\lambda_+ \lambda^{-1}_- s_1 t_2 \quad & 0 \quad   & - 2 s_1 t_2 \quad & 2 c_1 \quad & -\sqrt{2} \lambda_+ s_1 t_2 \\
			                   - \lambda_+ \lambda^{-1}_- c_1 t_2 \quad          & 0 \quad   & 2 c_1 t_2 \quad   & 2 s_1 \quad & \sqrt{2} \lambda_+ c_1 t_2  \\
			                   -2 \lambda_-^{-1} \quad                           & 0 \quad   & 0 \quad           & 0 \quad     & \sqrt{2}
		                   \end{pmatrix}
	\end{cases}
\end{equation}
for each sign choice and with the shorthand $s_i = \sin x^i$, $c_i = \cos x^i$, $ct_i = \cot x^i$, $t_i = \tan x^i$, and $\lambda_\pm = \lambda \pm 1$. The twisted mega-space torsion $T_{\Ah\Bh\Ch}$ is formed by the structure constants of SU(2)$\times$SU(2), namely
\begin{equation}
	T_{\Ah\Bh\Ch} = \llangle [ t_{\Ah}, t_{\Bh} ] , t_{\Ch} \rrangle\,,
\end{equation}
with $t_{\Ah} = \begin{pmatrix} t_a & t^a \end{pmatrix}$. As this group is semi-simple, its Lie algebra does not have any elements in its center. Hence, the Bianchi identity for the dilatonic twisted torsion requires $T_{\Ah} = 0$. Integrating \eqref{eqn:defThatA} we find
\begin{equation}
	\Phi = - \tfrac12 \log \left( c_2 s_2 \right)
\end{equation}
with might be shifted by an arbitrary constant. To obtain the physical fields of this background, we have to fix the generalized metric $\cH^{\Ac\Bc}$ such that it satisfies the constraint given in section~\ref{sec:GS}. A valid option is
\begin{equation}\label{eqn:cHexample}
	\cH^{\Ac\Bc} = \diag\left( 1, 1, 1, 1, \kappa_{\onei\onei} \right)\,.
\end{equation}
It can be used to read-off the metric, $B$-field and gauge potential from the explicit parameterization given in \eqref{eqn:cHparam}. Following this idea, we find
\begin{equation}\label{eqn:ds2}
	\dd s^2 = \frac{\sin^2( 2 x^2 )}{2 \left[ \lambda - \cos(2 x^2) \right]^2} (\dd x'^1)^2 + 2 (\dd x^2)^2\,,
\end{equation}
and the gauge potential
\begin{equation}
	A = \begin{cases}
		\frac{\sqrt{2} (\lambda-1) \cos^2 (x^2)}{\cos(2 x^2)-\lambda} \dd x'^1 \\
		\frac{\sqrt{2} (\lambda+1) \sin^2 (x^2)}{\cos(2 x^2)-\lambda} \dd x'^1
	\end{cases}
\end{equation}
with $x'^1 = (\lambda \pm 1) x^1$. The latter comes with the non-vanishing field strength while the $B$-field vanishes. We the metric, we can finally extract the dilaton
\begin{equation}\label{eqn:dilaton}
	\phi = \Phi + \tfrac12 \log \sqrt{g} = - \tfrac12 \log | \lambda - \cos(2 x^2) | + \phi_0\,.
\end{equation}

This background is closely related to the two-dimensional Euclidean black hole geometry discussed in \cite{Witten:1991yr}. After the analytic continuation
\begin{equation}
	x'^1 = 2 \theta \,, \qquad \text{and} \qquad x^2 = i r\,,
\end{equation}
the metric \eqref{eqn:ds2} and the dilaton \eqref{eqn:dilaton} become
\begin{align}
	\dd s^2 & = - 2 \left[ (\dd r)^2 + \frac{\sinh(2 r)^2}{[\lambda - \cosh(2r)]^2} (\dd \theta)^2 \right]\,,
	        &                                                                                                 & \text{and} &
	\phi    & = - \tfrac12 \log | \lambda - \cosh( 2 r ) | + \phi_0\,,
	\intertext{respectively. For $\lambda = - 1$, these expressions simplify to}\label{eqn:cigar}
	\dd s^2 & = - 2 \left[ (\dd r)^2 + \tanh^2 r (\dd \theta)^2 \right]\,,                                    &            & \text{and} &
	\phi    & = - \log \cosh r + \phi_0 \,,
	\intertext{(known as the cigar geometry), while for $\lambda = 1$, they become}\label{eqn:trumpet}
	\dd s^2 & = - 2 \left[ (\dd r)^2 + \coth^2 r (\dd \theta)^2 \right]\,,                                    &            & \text{and} & \phi & = - \log \sinh r + \phi_0 \,,
\end{align}
known as the T-dual trumpet geometry. For both cases, the field strength
\begin{equation}
	F = \pm 2 \sqrt{2} \frac{(\lambda^2 - 1) \sinh(2 r)}{[\lambda - \cosh(2 r)]^2} \dd \theta \wedge \dd r
\end{equation}
of $A$ vanishes. Hence, we found a background which interpolates between the cigar and the trumpet.

Both backgrounds, \eqref{eqn:cigar} and \eqref{eqn:trumpet} are captured by conformal field theories on the worldsheet. Therefore, their $\beta$-functions for the metric and $B$-field have to vanish and we expect the that the variation of the effective action \eqref{eqn:ShetDFT} with respect to these fields vanishes, too. This can be checked by computing the field equations for the action \eqref{eqn:ShetDFT} \cite{Geissbuhler:2013uka}. In doing so, we will find that the generalized fluxed $F_{\Ac\Bc\Cc}$ and $F_{\Ac}$ are not constant any more. This is not surprising, because we know from the analysis in section~\ref{sec:tgT} that only the combination in \eqref{eqn:THet},
\begin{equation}
	\THet_{\Ac\Bc\Cc} = \cF_{\Ac\Bc\Cc} - E_{[\Ac \deltai} (\tau^{\deltai})_{\Bc\Cc]} \,,
\end{equation}
will be constant. If condition \eqref{eqn:tauProductCanonical} holds the last term would vanish. But this condition is violated here. However, one can check that in this example the last term comes exclusively from the twist $\Mt$ in the mega-space generalized frame. The latter is by construction an element of the double Lorentz group and thus can be removed by an compensating double Lorentz transformation. There is only one challenge: These required transformation depends on the coordinates related to the gauge group $\GH$. Thus, it will violate the section condition \eqref{eqn:hetSC}. We will always encounter this phenomena when \eqref{eqn:tauProductCanonical} does not hold. Fortunately, it is possible to include the underlined terms of \eqref{eqn:ShetDFT} in the action and get the corresponding field equations as described in \cite{Geissbuhler:2013uka}. Because the twisted frame still satisfies the closure constraint explained there this approach is totally fine. And indeed, one finds this way that the equations of motion for the generalized dilaton are satisfied not only for $\lambda=\pm 1$ but for all values of $\lambda$. A check for this result is to repeat the computation with the original non-constant generalized fluxes. It leads to the same conclusion. The only non-vanishing field equation is the one for the generalized dilaton. It is proportional to the generalized Ricci scalar which due to $T_{\Ah}=0$ takes the simple form
\begin{equation}
	\mathcal{R} = - \tfrac14 \cH^{\Ac\Dc} \THet_{\Ac\Bc\Cc} \THet_{\Dc}{}^{\Bc\Cc} + \tfrac1{12} \cH^{\Ac\Bc} \cH^{\Cc\Dc} \cH^{\Ec\Fc} \THet_{\Ac\Cc\Ec} \THet_{\Bc\Dc\Fc} - \tfrac14 \cH^{\Ac\Dc} \THet_{\Ac\Bc}{}^{\gamman} (\htau_{\gamman})_{\Dc}{}^{\Bc}  = 2\,.
\end{equation}
Again, we see that it does not depend on the parameter $\lambda$. Moreover, $\mathcal{R}$ is constant which is expected from $\lambda = \pm 1$ because there it is proportional to the central charge.

We conclude that even the simplest conceivable example for the extension of the \PS{} construction we present in this article gives rise to interesting new heterotic backgrounds. A natural thing one might want to study next are admissible deformations of the generalized metric \eqref{eqn:cHexample}. For $\lambda=\pm 1$ they for example in include the $\lambda$ deformed two-sphere whose worldsheet theory is integrable.

\section{Conclusions}\label{sec:conclusions}
Starting point for this article has been the question if it is possible to drop the isotropy of the generalized structure group for the \PS{} construction. Considering the relation between the latter and symplectic reductions of Courant algebroids, it was expected that the construction should work even without this constraint and then should relate to heterotic/type I strings whose gauge structure is governed by transitive Courant algebroids. This expectation has been confirmed and we furthermore witnessed that a versatile structure emerges from an at the first glance quite innocent extension.

An important application of the presented construction is to obtain backgrounds which solve the field equations of the respective leading-order, low-energy effective actions as illustrated by the example in section~\ref{sec:cigarToTrumpet}. Because their underlying algebraic structure is manifest, the analysis of generalized dualities and consistent truncations for heterotic/type I strings, whose exploration only begun recently \cite{Hatsuda:2022zpi,Hassler:2023nht,Mori:2024sfg}, is simplified considerably. A distinguished subclass in the relevant family of generalized homogenous spaces leads to integrable $\sigma$-models. Again, they have been mostly studied for bosonic strings and type II superstrings while only few results have been reported in the heterotic sector. A major advantage in our approach is that their mega-space is always a generalized group manifold (generalized parallelizable space) for the duality group O($D$,$D$). These space have been extensively studied over the last years and we understand their structure very well. Going from O($d$, $d$) to O($d$, $d+q$) with gauge group $\GH$ comes with challenges like the requirement of section condition violating twists that have to be introduced by hand. From the new perspective we take here, all the additional ingredients arise automatically from the reduction of the mega-space.

A second driving force behind the presented results is the construction of higher-derivative corrections in generalized geometry. At the moment, the most sophisticated technique to do so is called the generalized Bergshoeff-de Roo (gBdR) identification \cite{Baron:2018lve,Baron:2020xel}. It builds on the original idea of Bergshoeff and de Roo \cite{Bergshoeff:1989de} to extend the gauge group of the heterotic string by the Lorentz group and then identify the additional connection components with the spin connection. Thereby parts of the kinetic term of the gauge field generate a Riemann square term. Remarkably, the gBdR identification generates not only the leading (4-derivative) corrections but an infinite tower of higher derivatives. Unfortunately at the moment, these still lack the well established geometrical interpretation of the leading order, ruling out any application which relies on them -- like generalized duality or consistent truncations. Moreover, it makes it very hard to understand the underlying reasons for recent no-go theorems suggesting that $\xi(3)$-corrections cannot be captured in generalized geometry \cite{Hsia:2024kpi}. The heterotic version of the \PS{} construction presented here, is the key to gaining a better understanding of the mathematical structures that underlie the gBdR identification \cite{Gitsis:2024}. From a more geometrical perspective in terms of torsion constraints and partial gauge fixing, both are concepts which already appear in general relativity and gauge theories.

\section*{Acknowledgements}
We would like to thank Daniel Butter, Achilleas Gitsis and Luca Scala for helpful discussions. The work of FH is supported by the SONATA BIS grant 2021/42/E/ST2/00304 from the National Science Centre (NCN), Poland. The research of DO~is part of the project No. 2022/45/P/ST2/03995 co-funded by the National Science Centre and the European Union’s Horizon 2020 research and innovation programme under the Marie Sk\l odowska-Curie grant agreement no. 945339. The work of YS is supported by JSPS KAKENHI Grant Number JP23K03391.

\vspace{10pt}
\includegraphics[width = 0.09 \textwidth]{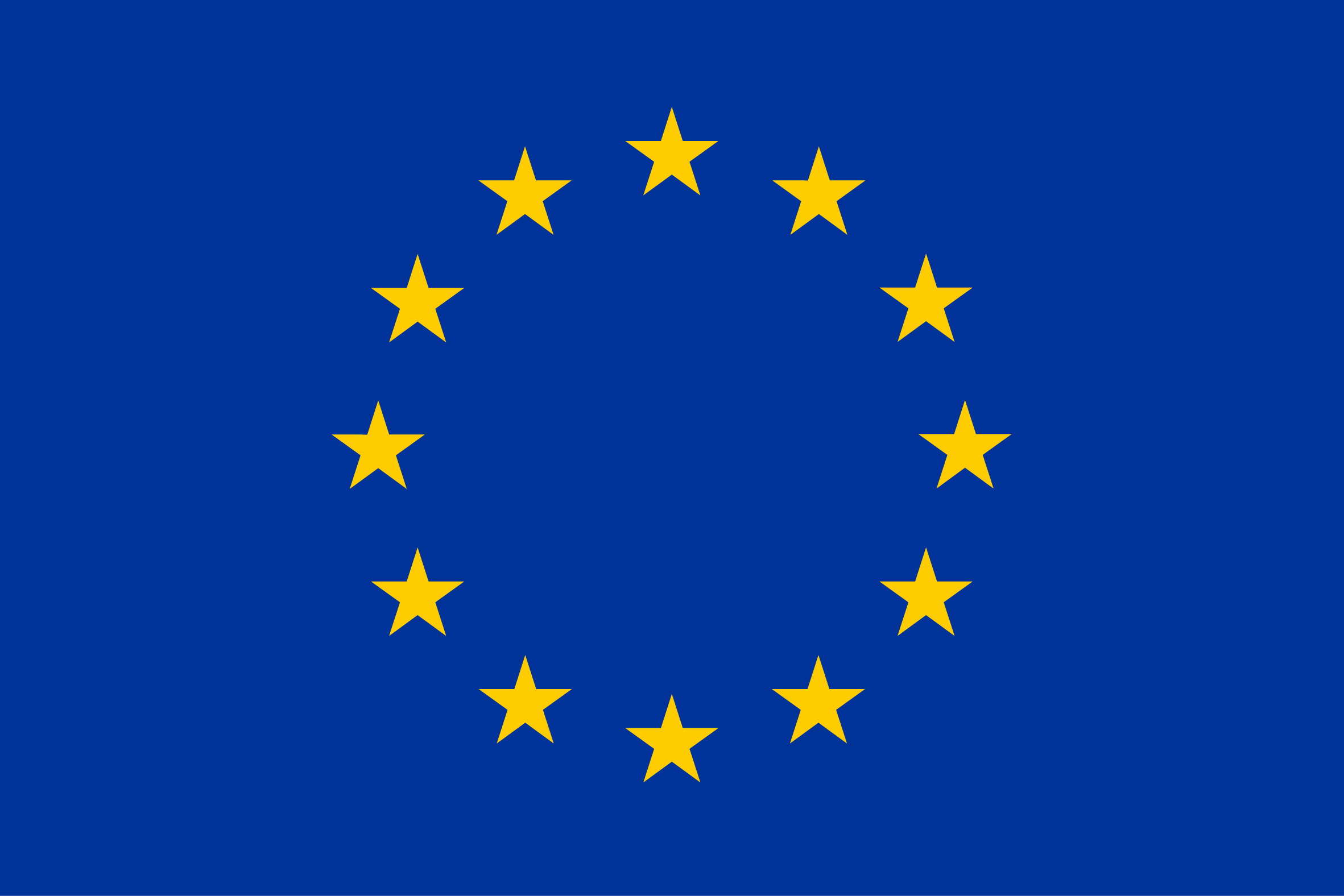} $\quad$
\includegraphics[width = 0.7 \textwidth]{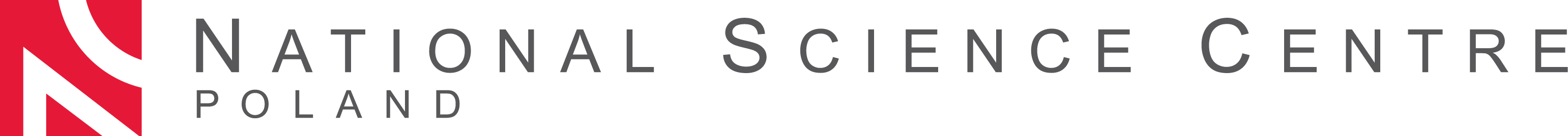}

\appendix
\section{Commutators}\label{app:commutators}
In this appendix, all non-vanishing commutation relations between the $\Odd[d+n]$ generators that are not present in the main text are included for completeness:
\begin{equation}
	\begin{aligned}
		\relax[\Rt^{\alpha}_A, K^{\gamma}_{\beta}] & = -\delta_{\beta}^{\alpha}\Rt^{\gamma}_A, \hspace{44.7mm} [\Rt^{\alpha\beta}, K^{\gamma}_{\delta}] = 2\delta_{\delta}^{[\alpha}\Rt^{\beta]\gamma},                                                                                                                                                                                                   \\
		[\Rt^{\alpha\beta}, R^B_{\gamma}]          & = 2\delta_{\gamma}^{[\alpha}\Rt^{\beta]}_C \eta^{C B}, \hspace{38mm} [\Rt^{\alpha}_A, R_{\beta\gamma}] = - 2\delta^{\alpha}_{[\beta}R^C_{\gamma]}\eta_{C A},                                                                                                                                                                                         \\
		[K_{\alpha}^{\beta}, R^C_{\gamma}]         & = -\delta_{\gamma}^{\beta}R^A_{\alpha} +\tfrac{1}{2}\kappa_{\alpha\gamma}\Rt^{\beta}_C +\tfrac{1}{2}\kappa_{\alpha\delta}\delta_{\gamma}^{\beta}\Rt_C^{\delta},                                                                                                                                                                                      \\
		[K_{\alpha}^{\beta}, R_{\gamma\delta}]     & = 2\delta^{\beta}_{[\gamma}R_{\delta]\alpha} -\kappa_{\alpha[\gamma}K_{\delta]}^{\beta}-\kappa_{\alpha\epsilon}\delta^{\beta}_{[\gamma}K_{\delta]}^{\epsilon} +\tfrac{1}{2}\kappa_{\alpha_{[\gamma}}\kappa_{\delta]\epsilon}\Rt^{\beta\epsilon}-\tfrac{1}{2}\kappa_{\alpha\epsilon}\kappa_{\zeta[\gamma}\delta_{\delta]}^{\beta}\Rt^{\epsilon\zeta}, \\
		[\Rt^{\alpha\beta}, R_{\gamma\delta}]      & = -4\delta^{[\alpha}_{[\gamma}K_{\delta]}^{\beta]} -2 \delta_{[\gamma|}^{[\alpha|}\kappa_{|\epsilon|\delta]}\Rt^{|\epsilon|\beta]}.
	\end{aligned}
\end{equation}
Commutators that neither appear here nor in the main text, such as $[K_{A B}, R_{\alpha \beta}]$ or $[\Rt^{\alpha\beta}, \Rt^{\gamma\delta}]$, are trivial.

\bibliography{literature}

\bibliographystyle{JHEP}

\end{document}